\documentclass[conference]{IEEEtran}
\pdfoutput=1
\usepackage{stfloats}
\usepackage{caption}
\usepackage{subcaption}
\usepackage{graphics}      %
\usepackage[T1]{fontenc}   %
\usepackage{diagbox}
\usepackage{mathptmx}
\usepackage{booktabs}
\usepackage{textcomp}
\usepackage{xspace}
\usepackage{lipsum}

\usepackage{epsfig,endnotes,makecell,array,multirow}
\usepackage{float}
\usepackage{xspace}
\usepackage{color}
\usepackage{csquotes}
\usepackage{amsmath, amsfonts}
\usepackage{booktabs}
\usepackage{mdwtab}
\usepackage{soul}
\usepackage{bbm}
\usepackage{hyperref}
\usepackage{tabularx}
\usepackage{hhline}
\usepackage{enumitem}

\DeclareMathOperator*{\argmin}{arg\,min}

\usepackage{microtype}        %
\usepackage[all]{hypcap}    %
\usepackage{ccicons}          %
\usepackage{todonotes}
\newcommand*{\etal}{{\em et al.}\@\xspace}
\newcommand*{\eg}{{\em e.g.,}\@\xspace}
\newcommand*{\ie}{{\em i.e.,}\@\xspace}
\usepackage{pifont}%

\usepackage{regexpatch}

\newcommand{\name}{{\fontfamily{cmss}\selectfont{Face-Off}}\xspace}
\newcommand\norm[1]{\left\lVert#1\right\rVert}

\DeclareMathOperator{\bfx}{{\bf x}}

\DeclareMathOperator{\bfa}{{\bf a}}

\makeatletter
\g@addto@macro\normalsize{%
\setlength\abovedisplayskip{2pt}
\setlength\belowdisplayskip{2pt}
\setlength\abovedisplayshortskip{3pt}
\setlength\belowdisplayshortskip{3pt}
}
\makeatother

\newcommand\blfootnote[1]{%
  \begingroup
  \renewcommand\thefootnote{}\footnote{#1}%
  \addtocounter{footnote}{-1}%
  \endgroup
}

\begin{document}

\title{\vspace*{-0.5in}
{{\normalsize \rm In 21\textsuperscript{st} {\em Proceedings of  Privacy Enhancing Technologies Symposium}\hrule}}
\vspace*{0.4in}Face-Off: Adversarial Face Obfuscation}

\author{Varun Chandrasekaran\text{*}, Chuhan Gao\text{*}\IEEEauthorrefmark{3}, Brian Tang, Kassem Fawaz, Somesh Jha, Suman Banerjee\vspace*{0.15cm} \\ 
Microsoft\IEEEauthorrefmark{3}, University of Wisconsin-Madison 
}

\maketitle

\begin{abstract}
Advances in deep learning have made face recognition technologies pervasive. While useful to social media platforms and users, this technology carries significant privacy threats. Coupled with the abundant information they have about users, service providers can associate users with social interactions, visited places, activities, and preferences--some of which the user may not want to share. Additionally, facial recognition models used by various agencies are trained by data scraped from social media platforms. Existing approaches to mitigate these privacy risks from unwanted face recognition result in an imbalanced privacy-utility trade-off to users. In this paper, we address this trade-off by proposing \name, a privacy-preserving framework that introduces strategic perturbations to the user's face to prevent it from being correctly recognized. To realize \name, we overcome a set of challenges related to the black-box nature of commercial face recognition services, and the scarcity of literature for adversarial attacks on metric networks. We implement and evaluate \name to find that it deceives three commercial face recognition services from Microsoft, Amazon, and Face++. Our user study with 423 participants further shows that the perturbations come at an acceptable cost for the users.
\end{abstract}

\blfootnote{\text{*}Authors contributed equally.}

\section{Introduction}
\label{sec:introduction}

Enabled by advances in deep learning, face recognition permeates several contexts, such as social media, online photo storage, and law enforcement~\cite{clearview}. Online platforms such as Facebook, Google, and Amazon provide their users with various services built atop face recognition, including automatic tagging and grouping of faces. Users share their photos with these platforms, which detect and recognize the faces present in these photos. Automated face recognition, however, poses significant privacy threats to users, induces bias, and violates legal frameworks~\cite{chi:2007,clearview}. These platforms operate proprietary (black-box) recognition models that allow for associating users with social interactions, visited places, activities, and preferences--some of which the user may not want to share.

\begin{figure}[t]%
\centering
\includegraphics[width=\columnwidth]{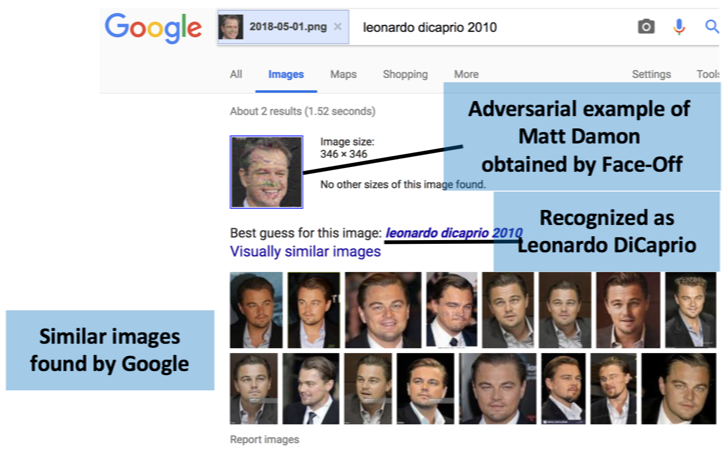}
\caption{Adversarial example of \texttt{Matt Damon} generated by \name recognized as \texttt{Leonardo DiCaprio} by Google image search.}
\label{fig:intro_pic_2}
\end{figure}

Existing approaches to mitigate these privacy risks result in an imbalanced trade-off between privacy and utility. Such approaches rely on (a) blurring/obscuring/morphing faces~\cite{gross2009face}, (b) having the users utilize physical objects, such as eyeglass frames, clothes, or surrounding scenery with special patterns~\cite{glass1,glass2}, and (c) evading the face detector (the necessary condition for face recognition)~\cite{bose2018adversarial}. These solutions, however, exhibit two main drawbacks to users. First, the user can no longer meet their original goal in using the social media platform, especially when various applications built atop of face detection (such as face-enhancement features) are broken. Second, specially manufactured objects for physical obfuscation are not omnipresent and might not be desirable by the user.

Relying on insights from previous work~\cite{glass1, glass2,bose2018adversarial}, we propose a new paradigm to improve the trade-off between privacy and utility for such users. In adversarial machine learning, carefully crafted images with small and human-imperceptible perturbations cause misclassifications~\cite{szegedy2013intriguing,goodfellow2014explaining,carlini2017towards}. In this paper, {\em we extend this approach from classification models to metric learning,} as used in face recognition systems. In particular, we propose \name, a system that preserves the user's privacy against {\em real-world} face recognition systems. By carefully designing the adversarial perturbation, \name targets only face recognition (and not face detection), preserving the user's original intention along with context associated with the image. \name detects a user's face from an image to-be-uploaded, applies the necessary adversarial perturbation, and returns the image with a perturbed face. However, the design of \name faces the following challenges:

\begin{itemize}
\itemsep0em
\item Unlike classification networks, metric learning networks (used for face verification/recognition) represent inputs as feature embeddings~\cite{facenet,deepface,sphereface}. Real-world face recognition maps the feature embedding of an input image to the closest cluster of faces. Existing approaches target classification networks and must be retrofit for metric learning. 

\item As the models used by these organizations are proprietary, \name needs to perform {\em black-box attacks}. This issue is already challenging in the classification domain~\cite{szegedy2013intriguing,carlini2017towards,papernot2016practical}. Further, \name cannot use the service provider's face recognition API as a black-box oracle to generate the adversarially perturbed face~\cite{glass1} for two reasons. First, generating an adversarial example requires querying the API extensively, which is not free and is often rate-limited\footnote{Approaches based on gradient-free optimization~\cite{zhao2019design} are prohibitively expensive.}. Second, querying the black-box model defeats our purpose for privacy protection as it sometimes begins with releasing the original face~\cite{glass1}.  
\end{itemize}

We address the first challenge by designing two new loss functions specifically targeting metric learning. These loss functions aim to pull the input face away from a cluster of faces belonging to the user (in the embedding space), which results in incorrect face recognition. Both loss functions can be integrated with the state-of-the-art adversarial attacks against classification networks~\cite{carlini2017towards,madry2017towards}.

To meet the second challenge, we leverage {\em transferability}, where an adversarial example generated for one model is effective against another model targeting the same problem. We rely on surrogate face recognition models (which we have full access to) to generate adversarial examples. Then, \name amplifies the obtained perturbation by a small multiplicative factor to enhance transferability. This property reduces the probability of metric learning networks correctly recognizing the perturbed faces. Further, we explore amplification, beyond classifiers~\cite{cao2017mitigating,liu2016delving}, and show it enhances transferability in metric learning and reduces attack run-time.

We evaluate \name across three major commercial face recognition services: Microsoft Azure Face API~\cite{azure}, AWS Rekognition~\cite{aws}, and Face++~\cite{facepp}. \name generates perturbed images that transfer to these three services, preventing them from correctly recognizing the input face. Our adversarial examples also transfer to Google image search (refer Figure~\ref{fig:intro_pic_2}) successfully with the target labels. Based on a longitudinal study ({\em across two years}), we observe that commercial APIs have not implemented defense mechanisms to safeguard against adversarial inputs. 
We show that using adversarial training~\cite{madry2017towards} as a defense mechanism deteriorates natural accuracy, dropping the accuracy by 11.91 percentage points for a subset of the VGGFace2 dataset.
Finally, we perform two user studies on Amazon Mechanical Turk with 423 participants to evaluate user perception of the perturbed faces. We find that users' privacy consciousness determines the degree of acceptable perturbation; privacy-conscious users are willing to tolerate greater perturbation levels for improved privacy. 

In summary, our contributions are:
\begin{enumerate}
\item We propose two new loss functions to generate adversarial examples for metric networks (\S~\ref{sec:attacks_metric_embeddings}). We also highlight how amplification improves transferability for metric networks (\S~\ref{sec:intuition}).
\item We design, implement, and evaluate \name, which applies adversarial perturbations to prevent real-world face recognition platforms from correctly tagging a user's face (\S~\ref{sec:evaluation}). We confirm \name's effectiveness across three major commercial face recognition services: Microsoft Azure Face API, AWS Rekognition, and Face++.
\item We perform two user studies (with 423 participants) to assess the user-perceived utility of the images that \name generates (\S~\ref{sec:user}).
\end{enumerate}

\section{Background}
\label{sec:background}

This section describes the machine learning (ML) notation required in this paper. We assume a data distribution
$D$ over $\mathbf{X} \times \mathbf{Y}$, where $\mathbf{X}$ is the sample space and $\mathbf{Y} = \{ y_1,\cdots,y_L \}$ is the finite space of labels. For example, $\mathbf{X}$ may be the space of all images, and $\mathbf{Y}$ may be the labels of the images.

\vspace{1mm}
\noindent{\bf Empirical Risk Minimization:} In the {\it empirical risk minimization (ERM)} framework, we wish to solve the following optimization problem:
\[
w^* = \min_{w \in H} \; \mathbb{E}_{(\bfx,y) \sim D} \; \mathcal{L}(w,\bfx,y),
\]
where $H$ is the hypothesis space and $\mathcal{L}$ is
the loss function (such as cross-entropy loss~\cite{zhang2018generalized}). We denote vectors in bold (\eg $\bfx$). Since the distribution is usually unknown, a learner solves
the following problem over a dataset $S_{train} = \{ (\bfx_1,y_1),
\cdots. (\bfx_n,y_n) \}$ sampled from $D$:
\[
w^* = \min_{ w \in H } \; \frac{1}{n} \sum_{i=1}^n  \mathcal{L}(w,\bfx_i,y_i)
\]
Once the learner has solved the optimization problem given above, it obtains a solution
$w^* \in H $ which yields a classifier $F: \mathbf{X} \rightarrow
\mathbf{Y}$ (the classifier is usually parameterized by $w^*$ \ie, $F_{w^*}$, but we will omit this dependence for brevity).

\subsection{Metric Embeddings}
\label{sec:def_metric_embeddings}

A {\it deep metric embedding} $f_\theta$ is function from $\mathbf{X}$ to $\mathbb{R}^m$, where
$\theta \in \Theta$ is a parameter chosen from a parameter space and $\mathbb{R}^m$ is the space of $m$-dimensional real vectors. Throughout the section, we 
sometimes refer to $f_\theta (\bfx)$ as the {\em embedding of $\bfx$}. Let $\phi: \mathbb{R}^m \times \mathbb{R}^m \rightarrow \mathbb{R}^+$ be a distance metric\footnote{For all definitions that follow, $\phi$ represents the 2-norm.} on $\mathbb{R}^m$. Given a metric embedding function $f_\theta$,
we define $d_f(\bfx, \bfx_1)$ to denote $\phi (f_\theta (\bfx), f_\theta (\bfx_1))$. $[n]$ denotes the set $\{1, \cdots, n\}$.

\vspace{2mm}
\noindent{\bf Loss Functions:} Deep embeddings use different loss functions than typical classification networks. We define two of these loss functions: contrastive and triplet.

\vspace{1mm}
For a constant $\gamma \in \mathbb{R}^+$, the {\it contrastive} loss is defined over the pair $(\bfx,y)$ and $(\bfx_1,y_1)$ of labeled samples from $\mathbf{X} \times \mathbf{Y}$ as:
$$
  \mathcal{L}(\theta,(\bfx,y),(\bfx_1,y_1)) =  \mathbb{I}_{y = y_1} \cdot d^2_f (\bfx,\bfx_1) + \mathbb{I}_{y \not= y_1} \cdot [\gamma - d^2_f (\bfx,\bfx_1)],
$$
where $\mathbb{I}_E$ is the indicator function for event
$E$ (and is equal to $1$ if event $E$ is true and $0$ otherwise).

\vspace{1mm}
The {\it triplet} loss is defined over three labeled samples--$(\bfx,y)$, $(\bfx_1,y)$ and $(\bfx_2,y_2)$, given a constant $\gamma \in \mathbb{R}^+$,  as:
$$  
  \mathcal{L}(\theta,(\bfx,y),(\bfx_1,y),(\bfx_2,y_2))  =  [ d_f^2 (\bfx,\bfx_1) - d_f^2 (\bfx,\bfx_2) + \gamma ]_+ ,
$$  
where $[x]_+=\max(x,0)$ and $y \neq y_2$. 

\vspace{1mm}
\noindent{\bf Inference:}
Let $A = \{ (\bfa_1,c_1),\cdots,
(\bfa_k,c_k)\}$ be a reference dataset (\eg a set of face and label pairs). Note that $A$ is the dataset used during inference  time and different from the dataset $S_{train}$ used for training. Let $A_y \subset A$ be the subset of the reference dataset with label $y$ (\ie $A_y = \{ (\bfa_j,y) \; \mid \; (\bfa_j,y) \in A \}$). 
Additionally, we denote the centroid of set $A_y$ as $\beta_{y,f} \in \mathbb{R}^m$. Formally, the centroid of label $y$ is defined as follows:
$$\beta_{y,f} = \frac{1}{|  A_y | } \; \sum_{ (\bfa_i,y) \in A_y} f_\theta(\bfa_i)$$

Suppose we have a sample $\bfx$ and a reference dataset $A$, let $$j^* = \argmin_{j \in [k]} \phi (\beta_{c_j,f},f_\theta(\bfx)).$$ We predict the label of $\bfx$ as $c_{j^*}$. 

\vspace{1mm}
\noindent{\bf Recognition vs. Matching:} In the face recognition setting, the training set $S_{train}$ corresponds to a large labeled dataset of individuals' faces. During inference, the face recognition service has access to a reference dataset $A$; these could correspond to tagged images on Facebook, for example. When a user uploads a new image, the service searches for the centroid that is closest to the image in the embedding space and returns the label corresponding to the centroid. In the face matching setting, the service provider receives two faces and returns the distance between them in the embedding space.

\subsection{Attacks on Metric Embeddings}
\label{sec:attacks_metric_embeddings}

\vspace{1mm}
\noindent{\bf Attack Overview:} We define two types of attacks on metric embedding networks: untargeted and targeted. In the formulations given above, we assume that $\mathbf{X}$ is a metric space with $\mu$ defined as a
metric on $\mathbf{X}$ (\eg $\mathbf{X}$ could be $\mathbb{R}^2$ with $\mu$ representing  $\ell_\infty$, $\ell_1$,
or $\ell_p$ norms for $p \geq 2$). $\delta$ is the perturbation we add to inputs from $\mathbf{X}$. We summarize the attacks below:

\vspace{1mm}
\noindent{\it 1. Untargeted attack} on $\bfx$ can be described as follows:
\[
\begin{array}{c}
  \min_{\delta \in \mathbf{X}} \; \mu (\delta) \;\\

  \mbox{\it s.t.} \; \argmin_{j \in [k]} \phi (\beta_{c_j,f},f_\theta(\bfx)) \neq  \argmin_{j \in [k]} \phi (\beta_{c_j,f},f_\theta(\bfx+\delta))
  
\end{array}
\]
This attack aims to find a small perturbation that pushes the perturbed example's embedding to a closer centroid than the original one.

\vspace{2mm}
\noindent{\it 2. Targeted attack} (with label $t \not= \argmin_{j \in [k]} \phi (\beta_{c_j,f},f_\theta(\bfx))$) can be described as follows:
\[
\begin{array}{c}
   \min_{\delta \in \mathbf{X}} \; \mu (\delta) \;\\
\mbox{\it s.t.} \; \argmin_{j \in [k]} \phi (\beta_{c_j,f},f_\theta(\bfx+\delta)) = t
\end{array}
\]
This attack aims to find a small perturbation that pushes the perturbed example's embedding to the centroid corresponding to a target label.

\vspace{1mm}
\noindent{\bf Approach Overview:} Let $\bfx$ be a sample that we wish to perturb.  Intuitively, an attack increases the distance between the perturbed sample's embedding (with $y$ as the true label) and that of all those other samples with label $y$. Empirically, we found this objective to be stronger than just pushing an embedding of a perturbed sample away from the centroid $\beta_{y,f}$. Next, for a deep embedding $f_\theta$ and set $A_y$, define

$$d'_{f}(\mathbf{z}, A_y) = \frac{1}{|A_y|} \sum_{(\mathbf{a}_i, y) \in A_y} d_f(\bfx, \mathbf{a}_i)$$

Observe that $d'_f(\bfx,A_y)$ is a differentiable function of $\bfx$, and thus prior work (\eg FGSM~\cite{szegedy2013intriguing}) can be used to generate the adversarial perturbation. 

\vspace{1mm}
\noindent{\bf Concrete Formulation:} For the untargeted case, we pose the attacker's optimization
problem as:
\[
\begin{array}{c}
  \max_{\delta \in \mathbf{X}} \; d'_f(\bfx+\delta,A_y) \;\\
  \mbox{\it s.t.} \; ||\delta||_p \leq \epsilon
\end{array}
\]

For targeted attacks, the adversary wishes to label the face as target $t$; we refer to the term $d'_f(\bfx+\delta,A_t)$ as the {\em target loss}. We define the following function $G(\bfx',t)$ (also known as {\em hinge loss}), where $\bfx' = \bfx + \delta$ is the perturbed sample as follows:
    $$G(\bfx+\delta,t)  =  [d'_f(\bfx+\delta,A_t) - 
    \max_{y \neq t}d'_f(\bfx+\delta,A_y) + \kappa]_{+}$$
where the margin $\kappa$ denotes the desired separation from the source label's samples. Once $G(\bfx',t)$ is defined, we can adapt existing algorithms, such as
Carlini \& Wagner (CW)~\cite{carlini2017towards}, to construct the perturbation. Thus, for targeted attacks, the adversary wishes solve the following optimization
problem:\\

\vspace{-2mm}
\[
\begin{array}{c}
  \min \; ||\delta||_p \;\\
  \mbox{\it s.t.} \; ||\delta||_p \leq \epsilon;\\
  G(\bfx + \delta, t) \leq 0
\end{array}
\]

\vspace{1mm}
\noindent{\bf Amplification:} We define amplifying a perturbation $\delta$ by $\alpha>1$ as scaling $\delta$ with $\alpha$. If the attack algorithm generates a perturbed sample $\bfx+\delta$, amplification returns $\bfx +\alpha\cdot\delta$.

\section{\name: Overview}
\label{sec:overview}

Here, we provide an overview of \name, which aims to preserve the user's visual privacy against social media platforms.

\subsection{System and Threat Models}
\label{sec:threat_model}

Online users upload photos of themselves (or others) to social media platforms (such as Facebook or Instagram). These platforms first utilize a {\em face detector}~\cite{feraund2001fast} to identify the faces in the photo, and then apply {\em face recognition}~\cite{turk1991face} to tag the faces. The face recognition module can employ (a) verification (\ie, determine whether an uploaded face matches a candidate person, tag, or label), or (b) top-1 matching (\ie, find the top candidate for a match given a set of candidates). In particular, these platforms have these two properties: 

\begin{enumerate}
\itemsep0em
\item They use {\em proprietary, black-box models} for face recognition. The models are trained on {\em private datasets} using architectures or parameters which are not public.
\item They can process user-uploaded images of {\em varying sizes, resolutions, and formats}. The platform (with high probability) recognizes faces in all of them.
\end{enumerate}

Upon tagging the people in the photo, the platform can perform additional inferences beyond the user's expectations~\cite{inference}. For example, the platform can infer the behavior of specific people, the places they visit, the activities they engage in, and their social circles~\cite{chi:2007}. Additionally, these {\em labeled} photos can be scraped by various services and later used by various governmental agencies~\cite{clearview}. The prolonged analysis of user-uploaded images allows the platform (and other agencies that use these photos) to profile users, which enables targeted advertising~\cite{ads}, and lays the foundations for surveillance at the behest of a nation-state~\cite{nation}. Thus, it is essential to safeguard the privacy of user-uploaded media, specifically images, from social network providers. 

\noindent{\bf Actors:} In such a setting, we assume that the social network provider is an adversarial entity. The provider will analyze face tags to infer more information about the user.

\subsection {High-level Operation}
\label{sec:high_level}

\begin{figure*}[t]%
\centering
\includegraphics[width=\textwidth]{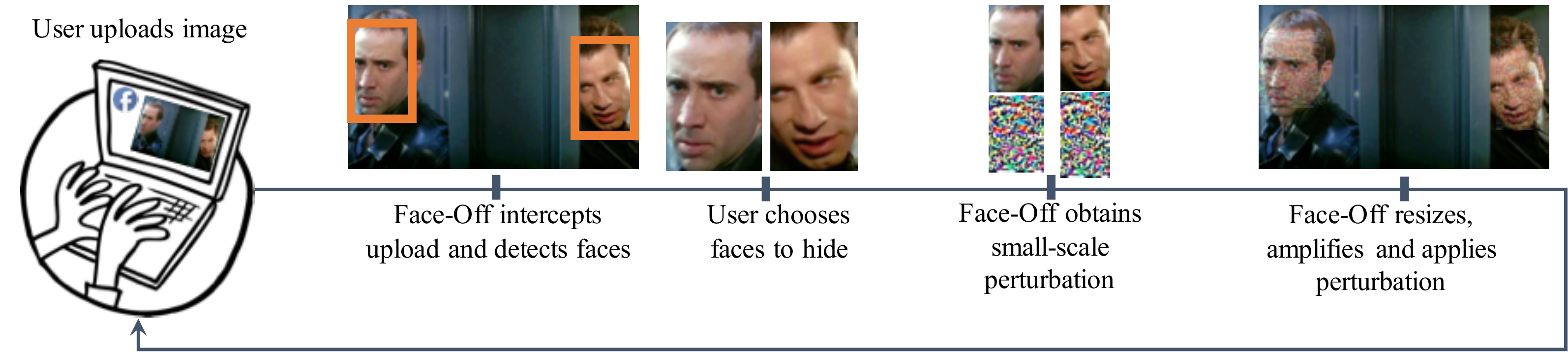}
\caption{High-level overview of \name's processing pipeline.}
\label{fig:high-level-fo}
\end{figure*}

\name aims to minimally modify images that the user wishes to upload such that the cloud provider cannot correctly recognize their face. Based on insights from adversarial ML (and specifically evasion attacks~\cite{biggio2013evasion} as highlighted in \S~\ref{sec:attacks_metric_embeddings}), \name applies small perturbation to the user inputs such that they are misclassified by facial recognition. \name sits between the user and the social network provider and generates pixel-level perturbations (or {\em masks}) to induce user-specified misclassifications. \name operates as follows (Figure~\ref{fig:high-level-fo}):

\begin{enumerate}
\itemsep0em
\item \name detects and extracts faces from user-uploaded photos. Advances in deep learning have made this process very accurate. In our implementation, we utilize an MTCNN~\cite{zhang2016joint}, which has a detection accuracy of 95\%.

\item Since our mask generation process requires inputs of a fixed size, \name resizes the detected faces, from the original size. This resizing process is also error-free.

\item \name proceeds to generate the mask for the resized face. We highlight the efficacy of this approach in \S~\ref{sec:evaluation}.

\item Finally, \name adds the generated mask to the resized face and returns a resized perturbed image. Sometimes, the generated masks are amplified by a  scalar constant $\alpha$. 
\end{enumerate}

Observe that, apart from step 3, {\em none of the other steps in the mask generation process induces erroneous artifacts in our pipeline}. Note that inaccuracies in detection do not lead to errors, but leads to the inability to generate a mask. In our threat model, users do not upload images that cause failures in the face detector~\cite{bose2018adversarial}. The procedure for mask generation (as required in step 3) always runs to completion, and we study the impact of various hyper-parameter choices on the success of the approach in \S~\ref{sec:offline}.

\vspace{1mm}
\noindent{\bf Challenges:} Achieving the functionality, as mentioned above, is challenging; we highlight several challenges below:

\begin{itemize}
\itemsep0em
\item [{\bf C1.}] Extensive work on evasion attacks (or generating adversarial examples) {\em focuses on classification}. However, \name requires attacks for metric embeddings (refer \S~\ref{sec:background}), which are {\em different}. Thus, new attack formulations, which include customized loss functions are required. 
\item [{\bf C2.}] Models used by the platform are black-box in nature, and there is a {\em lack of knowledge} of their internals. Most prior work on generating adversarial examples involves white-box access. To circumvent the issues associated with black-box access, we utilize {\em surrogate models} which we train. As these substitute models are similar to the proprietary models, we expect the generated adversarial examples to transfer~\cite{papernot2016transferability}. 
\item [{\bf C3.}] Since social media platforms are {\em capable of pre-processing} the images (through compression or resizing), it is essential that the perturbation generated for the scaled image transfers as well. 
\end{itemize}

In the rest of the paper, we highlight how \name overcomes these challenges. We reiterate that designing a system like \name is a challenging proposition because it provides privacy at inference time. Our work makes no assumptions about the nature of the models used by social network providers nor the data or mechanism they use for training. Adversarial knowledge and control of the model, the mechanism in which it is trained, and training data allow for different attack strategies (based on data poisoning~\cite{biggio2012poisoning}, for example). However, such threat models are unrealistic in practice.

\subsection{\name Design}
\label{subsec:implementation_surrogate}

\noindent{\bf Attacks:} We design \name by borrowing elements from two popular approaches: projected gradient descent (PGD) by Madry \etal~\cite{madry2017towards}, and the Carlini \& Wagner (CW) approach~\cite{carlini2017towards}. As noted in \S~\ref{sec:attacks_metric_embeddings}, we utilize two custom loss functions: (a) target loss and (b) hinge loss on surrogate model to which we have white-box access. Table~\ref{table:params} details the hyper-parameters of our attack implementations. In \S~\ref{sec:offline}, we describe in detail how (a) the choice of surrogate model, (b) choice of various attack hyper-parameters, and (c) choice of additional parameters such as amplification factor $\alpha$ and margin $\kappa$ impact attack success. All our code is available at \url{https://github.com/wi-pi/face-off}.

\begin{table}[t]
\begin{center}
\small
  \begin{tabular}{ c | c c c c}
    \toprule
    {\bf Parameters} & {\bf PGD} & {\bf CW$_{\infty}$} & {\bf CW$_{small}$} & {\bf CW$_{large}$} \\ 
    \midrule
    \midrule
    Perturbation Bound ($\varepsilon$) & 0.1 & - & - & - \\
    Norm ($p$) & 2 & $\infty$ & 2 & 2 \\
    Iterations ($N$) & 20/200 & 100 & 100 & 800\\
    Search Steps & - & 10 & 8 & 15\\
    Learning Rate ($\eta$) & 0.1 & 0.1 & 0.1 & 0.1 \\
    Initial Const. & 0.3 & 0.3 & 0.3 & 0.3 \\
    \midrule
    Hinge Loss & \checkmark & \checkmark& \checkmark& \checkmark\\
    Target Loss & \checkmark& \checkmark& \checkmark& \checkmark\\
    \bottomrule
    \end{tabular}
\end{center}
\caption{Attack hyper-parameters \vspace{-6mm}}
\label{table:params}
\end{table}

\vspace{1mm}
\noindent{\bf Surrogate Models:} We utilize two state-of-the-art face verification architectures: (a) the triplet loss architecture (\ie FaceNet~\cite{facenet}), and (b) center loss architecture (henceforth referred to as CenterNet~\cite{centerloss}). For both architectures, we utilize the code and pre-trained weights from the original repositories, and convert all implementations to \texttt{keras}~\cite{keras} to ensure compatibility with our perturbation generation framework (which was built using \texttt{tensorflow}~\cite{tf} and the \texttt{cleverhans} library~\cite{cleverhans}). The original implementations (collectively referred to as {\em small} models) accept inputs of the shape $96 \times 96 \times 3$ and $112 \times 96 \times 3$ respectively. We trained another variant of both these models (collectively referred to as {\em large} models), using the procedures outlined in the original papers\footnote{FaceNet (large) was obtained from the official \texttt{github} repository, while CenterNet (large) was trained from scratch.}, to accept inputs of shape $160 \times 160 \times 3$. Salient features, including test accuracy on the Labeled Faces in the Wild (LFW) dataset~\cite{huang2008labeled}, of these models are in Table~\ref{table:white}. In all models, the 2-norm between embeddings is used as the distance function $\phi(.,.)$.

\begin{table*}[t]
\begin{center}
  \begin{tabular}{c | c c c c c c}
    \toprule
    {\bf Abbreviation} & {\bf Architecture} & {\bf Dataset} & {\bf Loss} & {\bf Input Shape} & {\bf Embedding Size} & {\bf Test Accuracy}\\ 
    \midrule
    \midrule
    FSVT & FaceNet & VGGFace2 & Triplet & $96 \times 96 \times 3$ & 128 & 99.65 \%\\
    CSVC & CenterNet & VGGFace2 & Center & $112 \times 96 \times 3$ & 512 & 99.28 \%\\
    FLVT & FaceNet & VGGFace2 & Triplet & $160 \times 160 \times 3$ & 128 & 99.65 \%\\
    FLVC & FaceNet & VGGFace2 & Center & $160 \times 160 \times 3$ & 512 & 98.35 \%\\
    FLCT & FaceNet & CASIA & Triplet& $160 \times 160 \times 3$ & 512 & 99.05 \%\\
    \bottomrule
    \end{tabular}
\end{center}
\caption{Salient features of the white-box models used for offline mask generation.}
\label{table:white}
\end{table*}

\subsection{Theoretical Intuition}
\label{sec:intuition}
 
We discuss our intuition for why \name is effective in the untargeted attack case; an extension to the targeted case is trivial. Let $f: \mathbf{X} \rightarrow \mathbb{R}^m $ be the surrogate embedding (\eg generated by one of the models in Table~\ref{table:white}) and $g: \mathbf{X} \rightarrow \mathbb{R}^m $ be the victim embedding (\eg generated by an online model).  Recall that $\delta \in \mathbf{X}$ is the output of the untargeted attack algorithm (as defined in \S~\ref{sec:attacks_metric_embeddings}) that perturbs $\bfx \in \mathbf{X}$; the attack algorithm uses the surrogate embedding $f$. We consider the setting where the input to the attack is the sample $\bfx$ with a label $s$; both $f$ and $g$ label $\bfx$ as $s$, where $s=c_{j^*}$ such that $j^* =  \argmin_{j \in [k]} \phi (\beta_{c_j,f},f(\bfx))$ and $j^*= \argmin_{j \in [k]} \phi (\beta_{c_j,g},g(\bfx))$ (\ie inputs produce the same label using both metric learning networks)--as defined in \S~\ref{sec:def_metric_embeddings}. Define
the following variable:
\begin{eqnarray*}
  r(\bfx,\alpha,f, s) = \phi(f(\bfx + \alpha \cdot \delta),\beta_{s,f}) - \phi(f(\bfx),\beta_{s,f}).
\end{eqnarray*}

\vspace{1mm}
$r(\bfx,\alpha,f, s)$ denotes the change in the distance in the embedding space of $f(\bfx)$ from the centroid $\beta_{s,f}$ when we add the adversarial perturbation $\delta$ amplified by $\alpha \geq 1$. Our intuition is that $r(\bfx,\alpha,f, s)$ grows with the amplifcation factor $\alpha$. 

\begin{figure}[t]%
\centering
\includegraphics[width=0.8\columnwidth]{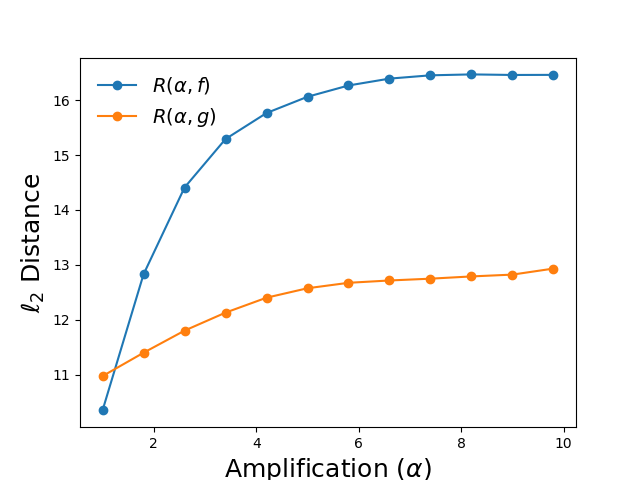}
\caption{The relationship between $R(\alpha,f)$ and $R(\alpha,g)$; amplifying the perturbation increases both terms.}
\label{fig:theoretical_intuition}
\end{figure}

We define $R(\alpha,f)$ as the expectation of $r(\bfx,\alpha,f,s)$:
\begin{eqnarray*}
  R(\alpha,f) = \mathbb{E}_{\bfx \sim D_{\mathbf{X}}} [r(\bfx,\alpha,f,s)]
\end{eqnarray*}
We empirically validate our intuition in Figure~\ref{fig:theoretical_intuition}, where amplification increases the value of $R(\alpha,f)$.

\vspace{1mm}
\noindent{\bf Embedding Similarity:} We state that two embeddings $f$ and $g$ are similar if the following holds:
\begin{eqnarray*}
  \forall \bfx \in \mathbb{R}^n ,\; \phi( g(\bfx) , f(\bfx)) \leq \omega(\bfx).
\end{eqnarray*}

Then, using the triangle inequality on the metric $\phi(\cdot,\cdot)$, it is clear to see that:
\begin{eqnarray*}
  r(\bfx,\alpha,g,s) \geq r(\bfx,\alpha,f,s) - 4 \omega(\bfx)
\end{eqnarray*}

In other words, if $r(\bfx,\alpha,f,s) > 4 \omega(\bfx)$, then $r(\bfx,\alpha,g,s) > 0$. This implies that for embedding $g$, $g(\bfx+ \alpha \cdot \delta$) is farther from the centroid $\beta_{s,g}$ than $g(\bfx)$, meaning the attack transfers to the victim model.  Taking the expectation of both the sides in the equation given above we get
\begin{eqnarray*}
  R(\alpha,g) \geq R(\alpha,f) - 4 \mathbb{E}_{\bfx \sim D}[ \omega(\bfx)]
\end{eqnarray*}
In particular, if $\omega(\bfx)$ is bounded by $\Delta$, we obtain the following:
\begin{eqnarray*}
  R(\alpha,g) \geq R(\alpha,f) - \Delta
\end{eqnarray*}

We empirically validate the claims above using the FLVT model as $f$ and the CSVC model as $g$ (refer Table~\ref{table:white}). Figure~\ref{fig:theoretical_intuition} reports $R(\alpha,f)$ and $R(\alpha,g)$. The values of $R(\alpha,f)$ and $R(\alpha,g)$ are averaged over 5 input samples. It is evident from the plot that amplification makes the perturbation {\em more} adversarial on both the surrogate and victim models.

\section{Parameter Choices}
\label{sec:offline}

\begin{figure*}[t]%
\centering
    \begin{subfigure}{0.24\textwidth}
        \centering
        \includegraphics[width=\textwidth]{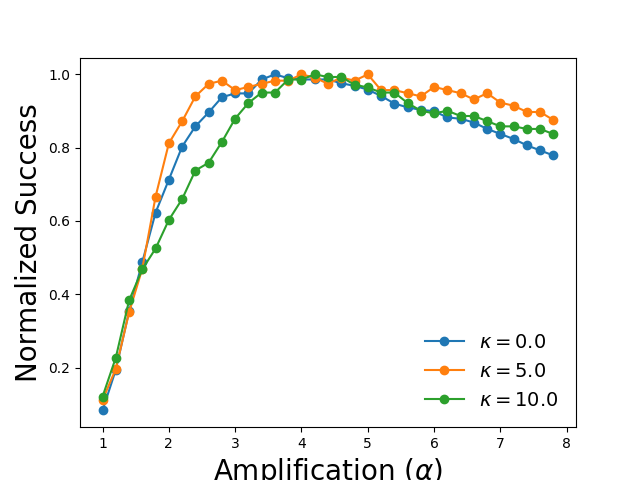}
        \caption{CW$_{\infty}$}
        \vspace{0.1in}
        \label{fig:whitebox_cwli_t}%
    \end{subfigure}
\begin{subfigure}{0.24\textwidth}
        \centering
        \includegraphics[width=\textwidth]{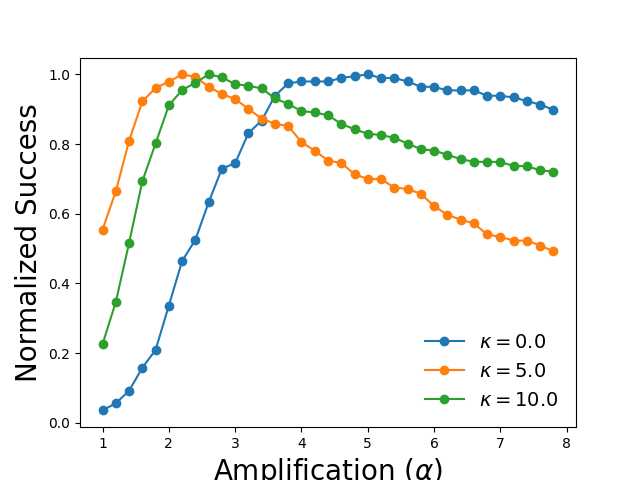}
        \caption{CW$_{small}$}
        \vspace{0.1in}
        \label{fig:whitebox_cwl2coarse_t}%
    \end{subfigure}
\begin{subfigure}{0.24\textwidth}
        \centering
        \includegraphics[width=\textwidth]{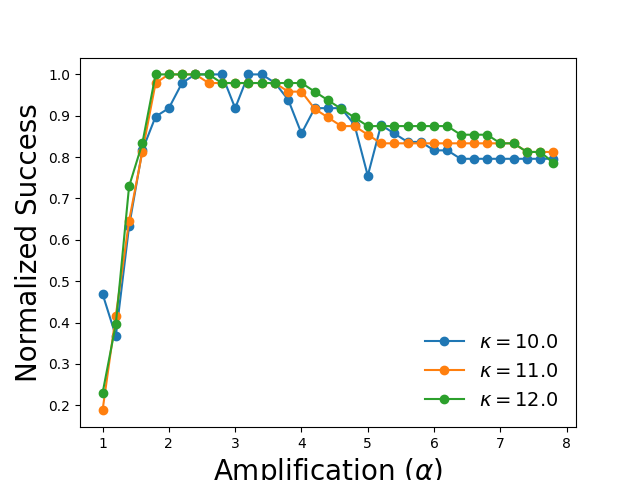}
        \caption{CW$_{large}$}
        \vspace{0.1in}
        \label{fig:whitebox_cwl2fine_t}%
    \end{subfigure}
\begin{subfigure}{0.24\textwidth}
        \centering
    \includegraphics[width=\textwidth]{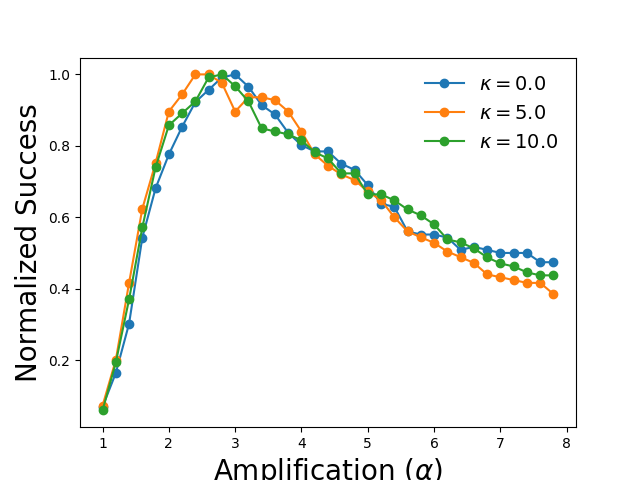}
        \caption{PGD}%
        \vspace{0.1in}
        \label{fig:whitebox_pgd_t}%
    \end{subfigure}
    \caption{White-box results for the \emph{targeted} top-1 attack with the FLVT model as the surrogate and the FLCT model as the victim.}
    \label{fig:whitebox_targeted}
\end{figure*}

\begin{figure*}[t]%
\centering
    \begin{subfigure}{0.24\textwidth}
        \centering
        \includegraphics[width=\textwidth]{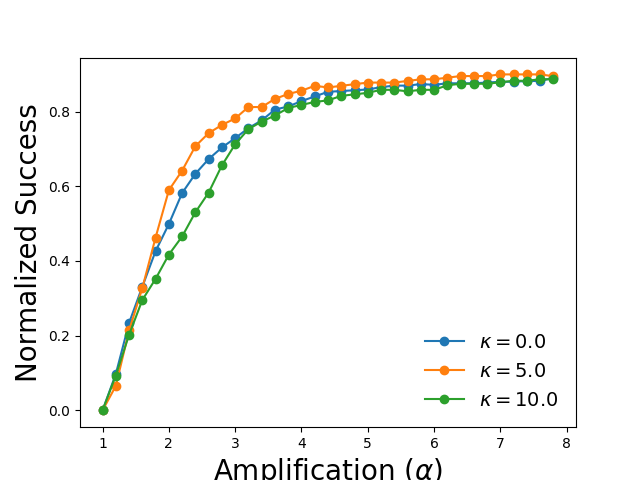}
        \caption{CW$_{\infty}$}
        \vspace{0.1in}
        \label{fig:whitebox_cwli_u}%
    \end{subfigure}
\begin{subfigure}{0.24\textwidth}
        \centering
        \includegraphics[width=\textwidth]{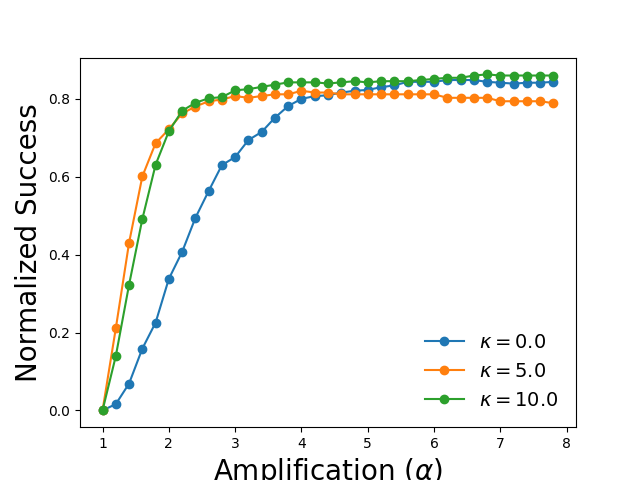}
        \caption{CW$_{small}$}
        \vspace{0.1in}
        \label{fig:whitebox_cwl2coarse_u}%
    \end{subfigure}
\begin{subfigure}{0.24\textwidth}
        \centering
        \includegraphics[width=\textwidth]{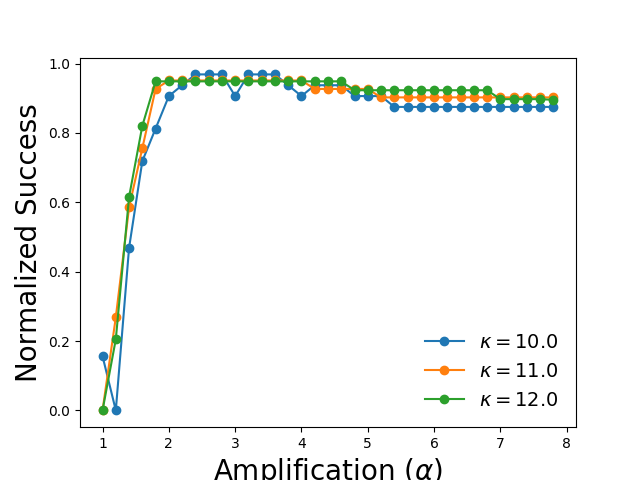}
        \caption{CW$_{large}$}
        \vspace{0.1in}
        \label{fig:whitebox_cwl2fine_u}%
    \end{subfigure}
\begin{subfigure}{0.24\textwidth}
        \centering
    \includegraphics[width=\textwidth]{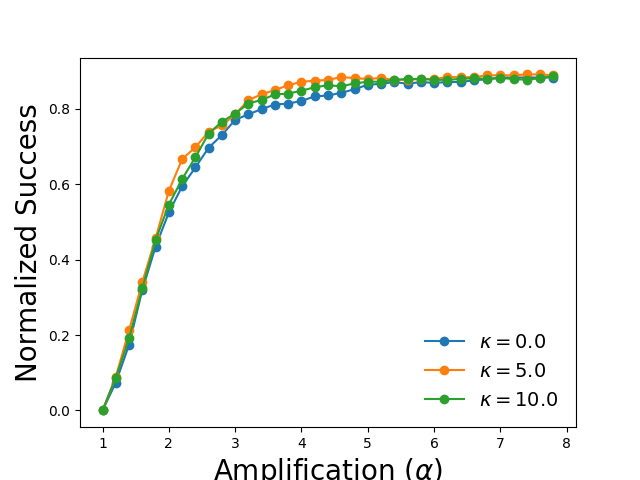}
        \caption{PGD}%
        \vspace{0.1in}
        \label{fig:whitebox_pgd_u}%
    \end{subfigure}
    \caption{White-box results for the \emph{untargeted} top-1 attack with the FLVT model as the surrogate and the FLCT model as the victim.}
    \label{fig:whitebox_untargeted}
\end{figure*}

Recall that our objective is to (a) generate adversarial examples (or masked samples) on a local surrogate model to which we have white-box access, and (b) transfer these examples to the black-box victim models used by social network providers. We conduct a comprehensive analysis to understand black-box transferability for both targeted (Figure~\ref{fig:whitebox_targeted}) and untargeted attacks (Figure~\ref{fig:whitebox_untargeted}). 

\vspace{1mm}
\noindent{\bf Setup:} We utilize images of celebrities, including some diversity in age, gender, and race. They are \texttt{Barack Obama, Bill Gates, Jennifer Lawrence, Leonardo DiCaprio, Mark Zuckerberg, Matt Damon, Melania Trump, Meryl Streep, Morgan Freeman}, and \texttt{Taylor Swift}. Thus, our experiments include images from 10 labels (totaling 90 source-target pairs) for {\em portrait} images alone. We consider 2 models (CSVC, FLVT from Table~\ref{table:white}) to generate the adversarial examples using 2 different adversarial loss functions (target and hinge) for 3 attacks ($\ell_2$ and $\ell_{\infty}$ variants of CW and an $\ell_2$ variant of PGD, as defined in \S~\ref{sec:attacks_metric_embeddings}). We evaluate 6 choices of margin $\kappa$ (\ie $\kappa=0, 5, 10, 11, 12, 13$), along with 40 choices of amplification factor $\alpha$ (\ie $\alpha \in [1,8]$ at intervals of 0.2), running for 2 settings (few and many) in terms of the number of execution iterations $N$. The remaining hyper-parameters are as specified in Table~\ref{table:params}. Due to space constraints, we only report the results using the FLVT model as the surrogate and the FLCT model as the victim. Results from other model combinations show similar trends.

\vspace{1mm}
\noindent{\bf Metrics:} We measure the top-1 matching accuracy \ie given a set of candidate labels (all 10 in our case); a correct match is one where the distance to the correct label is the smallest (in the embedding space). For targeted attacks, we define the {\em success metric} (a value that lies in $[0,1]$) as the ratio between the number of times an adversarial example matches the intended target (\ie attack success) and the number of tests (\ie number of attacks). The larger the {\em success metric}, the more effective is the attack. For untargeted attacks, we define the {\em success metric} as one minus the ratio of the number of times the label of the adversarial example is the true (source) label and the number of adversarial samples created. Again, the higher the success metric, the more successful is the attack. We detail the conclusions obtained from our ablation study below. To measure the dependence of one factor (say $\alpha$ or $\kappa$) on the success metric, we keep all other parameters fixed (unless explicitly stated otherwise) as specified in Table~\ref{table:params}.

\vspace{1mm}
\noindent{\bf Description of Plots:} Figures~\ref{fig:whitebox_targeted} and~\ref{fig:whitebox_untargeted} highlight the impact of amplification ($\alpha$) and margin ($\kappa$) on the success metric. Each point in the plot is the average of the success metric across all 90 source-target pairs used.

\subsection{Choice of Attack}

Our analysis clearly shows that the exact choice of attack (CW or PGD) does not significantly impact transferability. Figures~\ref{fig:whitebox_cwl2coarse_t} and~\ref{fig:whitebox_pgd_t} (as well as Figures~\ref{fig:whitebox_cwl2coarse_u} and~\ref{fig:whitebox_pgd_u}) show that, for a given pair of surrogate and victim models, the success metric is relatively comparable across attacks (in both the targeted and untargeted setting). The minor variations can be attributed to variations in hyper-parameters as specified in Table~\ref{table:params}. This suggests that the loss functions proposed in \S~\ref{sec:attacks_metric_embeddings} enable successful transferability of generated adversarial examples (more than the exact attack). 

\noindent{\bf $\ell_2$ vs. $\ell_{\infty}$ attack:} Given a fixed set of other execution parameters, the exact choice of norm does not impact attack success (as witnessed in Figure~\ref{fig:whitebox_cwli_t} and~\ref{fig:whitebox_cwli_u}). We conducted an (IRB approved) user study involving 50 participants (each shown 20 pairs of $\ell_{\infty}$ and $\ell_2$ variants of CW-based masked samples) to determine if one type of attack was more favorable (due to the imperceptibility of the perturbation). The users were nearly undecided between the two conditions: out of the 1000 assessments, 468 favored the $\ell_{\infty}$ attack and 532 favored $\ell_2$ attack. We could not reject the null hypothesis that both conditions are equally favorable to the users ($p=0.15$).

\vspace{1mm}
\noindent{\bf Takeaway:} Exact choice of attack or norm does not (greatly) influence transferability or perceptibility.

\subsection{Amplification \& Margin}
\label{sec:amp_margin}

Across all attacks, we observe that success is directly correlated with increasing $\kappa$ and $\alpha$. This result holds regardless of the target model (upon which transferability is being measured), and is independent of the exact loss function used. It is, however, crucial to understand the difference between $\kappa$ and $\alpha$. The choice of $\kappa$ has a direct impact on the run-time of the approach. However, amplification is post-processing applied to the generated images (and is off the critical path). Thus, one can suitably compensate for low $\kappa$ by increasing $\alpha$, and improve the run-time of the approach. From our results in Figures~\ref{fig:whitebox_targeted} and~\ref{fig:whitebox_untargeted}, we observe that $\alpha$ more directly influences the transferability in comparison to $\kappa$. %
In particular, we observe that values of $\kappa \geq 5$ and $\alpha \geq 2$ are ideal for transferability. 

\vspace{1mm}
\noindent{\bf Takeaway:} $\alpha$ influences transferability more than $\kappa$. Additionally, smaller values of $\kappa$ are preferred to reduce the run-time.

\subsection{Number of Iterations}

We now focus on the number of iterations $N$ and its impact on success. We only consider the CW attack for the 2-norm. We plot the impact of $\kappa$ and $\alpha$ on success across two trails: the first with a fewer number of iterations ($N=100$) in Figures~\ref{fig:whitebox_cwl2coarse_t} and~\ref{fig:whitebox_cwl2coarse_u}, and the second with more iterations ($N=800$) in Figures~\ref{fig:whitebox_cwl2fine_t} and~\ref{fig:whitebox_cwl2fine_u}. We observe that increasing $N$ results in increased transferability (and this holds with increasing $\kappa$ and $\alpha$). We note, however, that increasing $N$ is a time-consuming process as it lies on the critical path. 

\vspace{1mm}
\noindent{\bf Takeaway:} Although increasing $N$ increases the run-time, it improves transferability.

\section{Evaluation Setup}
\label{sec:implementation}

In \S~\ref{subsec:implementation_surrogate}, we detail how to construct the surrogate models and other attack details. In this section, we describe the real-world victim models (\S~\ref{subsec:models}) and our evaluation setup (\S~\ref{sec:setup}).

\subsection{Victim Models}
\label{subsec:models}

We evaluate \name using 3 popular commercial recognition APIs: (a) Azure Face API~\cite{azure}, (b) Face++~\cite{facepp}, and (c) Amazon Rekognition~\cite{aws}. These APIs accept two images as input (henceforth referred to as one query). They return a confidence value indicating how similar these images are, and a matching threshold $\tau$ (the two images correspond to the same face when the confidence value is above the threshold). The salient features of these APIs are available in Table~\ref{table:black_box}. To ensure a consistent comparison, all confidence scores are normalized to values in $[0,1]$, and the threshold $\tau$ was chosen as $0.5$.

\begin{table}[t]
\begin{center}
  \begin{tabular}{c c c c}
    \toprule
    {\bf API} & {\bf Confidence} & {\bf Threshold ($\tau$)} & {\bf Cost}\\ 
    \midrule
    \midrule
    Azure Face & $[0,1]$ & 0.5 & \$0.001 \\
    Rekognition & $[0,100]$ & 50 & \$0.001 \\
    Face++ & $[0,100]$ & Dynamic & Free \\
    \bottomrule
    \end{tabular}
\end{center}
\caption{Online API black-box models}
\label{table:black_box}
\end{table}

\subsection{Experimental Setup}
\label{sec:setup}

All our experiments were carried out on a server with 2 Titan XPs and 1 Quadro P6000 GPUs. This server had 40 CPU cores, 125 GB of memory, and ran Ubuntu version 16.04 LTS.

\vspace{1mm}
\noindent{\bf 1. Requirements:} We choose to generate adversarial examples for celebrity faces (as in \S~\ref{sec:offline}) due to their vast availability on the public internet. For each celebrity image $\bfx$ (whose label is referred to as the {\em source} label $s$), we need to obtain (a) a corresponding {\em target} label $t$, (b) a set of source images to calculate $d'_f(\bfx, A_{s})$, and (c) a set of target images to calculate $d'_f(\bfx, A_{t})$. These are required by our loss formulations for the optimization procedure detailed in \S~\ref{sec:attacks_metric_embeddings}.

\vspace{1mm}
\noindent{\bf 2. Processing Pipeline:} 
The processing pipeline is detailed in \S~\ref{sec:high_level}. The cropped faces are downsized accordingly using bilinear interpolation of \texttt{OpenCV}\footnote{https://docs.opencv.org/2.4}. The downsized faces are used to obtain adversarial perturbations. We refer to these perturbed images as {\em cropped} images. \name extracts the perturbation mask, resizes it, and applies it to the original subject after amplification. We refer to these perturbed images (after upsizing) as {\em uncropped} images.  

\vspace{1mm}
\noindent{\bf 3. Measuring Transferability:} A successful attack is one where the sample is misclassified based on the top-1 accuracy. This is the case since we compare the adversarial image with the source image (and the original label) from which it was generated. In \S~\ref{sec:topn}, we discuss top-n matching accuracy for the targeted attacks. 

\vspace{1mm}
\noindent{\bf 4. Understanding the Plots:} For all results presented in \S~\ref{sec:evaluation}, we plot the transferability (measured by the confidence value returned by the corresponding APIs) against the  norm of the perturbation (\ie $||\alpha \cdot \delta||_{p=2, \infty}$) for varying values of $\kappa$ and $\varepsilon$ (as in the case of PGD). We intend to highlight how the perceptibility of the perturbation (observed with increasing norm) impacts transferability. We also assess if both cropped uncropped faces transfer to the victim model. For brevity, we only plot the results using uncropped images. Results related to cropped images can be found in Appendix~\ref{app:cropped}.

\section{Evaluation}
\label{sec:evaluation}

Our evaluation is designed to answer the following questions (and we provide our findings as responses to the questions):

\begin{enumerate}
\itemsep0em

\item \textbf{Do the generated adversarial examples transfer to commercial black-box APIs specified in \S~\ref{subsec:models}?} 

Using the 2-norm variants of PGD and CW, we are able to successfully transfer the generated adversarial examples to all 3 commercial APIs (refer \S~\ref{subsec:commercial}).

\item \textbf{Can commercial APIs deploy defense mechanisms (such as adversarial training) to safeguard themselves against masked samples?}

Based on a longitudinal study ({\em across two years}), we observe that commercial APIs have not implemented defense mechanisms to safeguard against adversarial inputs. We also observe that adversarial training induces a substantial decrease in recognition accuracy (\S~\ref{subsec:robustness}).
Further, we show that using adversarial training~\cite{madry2017towards} as a defense mechanism deteriorates natural accuracy. When evaluation is performed using a subset of VGGFace2, accuracy drops by 11.91 percentage points (Table~\ref{table:advtop}).

\item \textbf{If transferability is measured by the top-n accuracy instead of the top-1 accuracy, how effective are the generated masked samples?}

We observe that, for targeted masked samples, the top-3 accuracy is higher than the top-1 accuracy. However, increasing $\alpha$ decreases this value as well (\S~\ref{sec:topn}).

\end{enumerate}

\subsection{Transferability to Commercial APIs}
\label{subsec:commercial}

Due to space constraints, we report results only for the $\ell_2$ variants of CW and PGD and for the uncropped setting; the cropped setting displays similar trends as evident from Appendix~\ref{app:cropped}. We use the CSVC model (Table~\ref{table:white}) as the surrogate to generate the adversarial samples. We fix the number of iterations to 500 iterations for the CW attack and 50 iterations for the PGD attack; other hyper-parameters are in Table~\ref{table:params}.  The experimental parameters are as follows: we varied $\kappa$ (for CW) and $\varepsilon$ (for PGD) in $[0,5.8]$ at increments of 0.2, and varied $\alpha$ from $[1,5]$ at increments of 0.1. The results are obtained using images of \texttt{Matt Damon} as the source and \texttt{Leonardo Di Caprio} as the target, and all experiments were carried out in August 2018. Across all plots, lower confidence values correspond to better privacy gains.

\begin{figure*}[]%
\centering
	\begin{subfigure}{0.31\textwidth}
		\centering
		\includegraphics[width=0.83\textwidth]{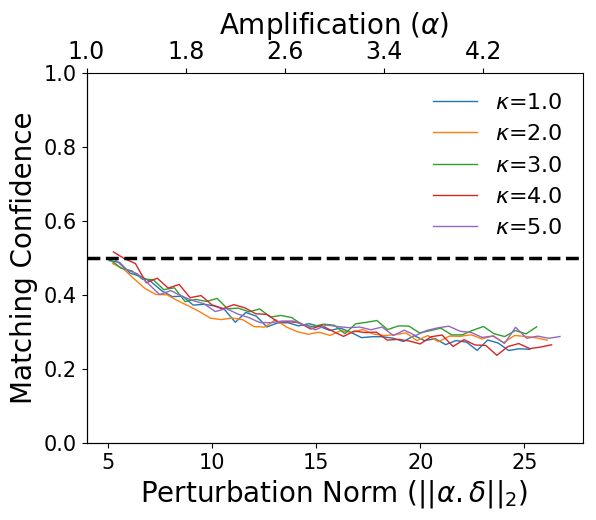}
        \caption{Azure}
        \vspace{0.1in}
		\label{fig:azure_cw_rr}%
	\end{subfigure}
	\hfill
    \begin{subfigure}{0.31\textwidth}
		\centering
		\includegraphics[width=0.83\textwidth]{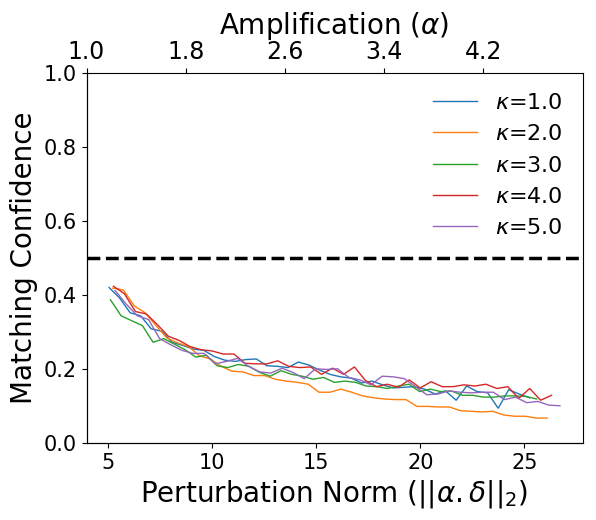}
		\caption{AWS Rekognition}
		\vspace{0.1in}
		\label{fig:awsverify_cw_rr}%
	\end{subfigure}
	\hfill
    \begin{subfigure}{0.31\textwidth}
		\centering
		\includegraphics[width=0.83\textwidth]{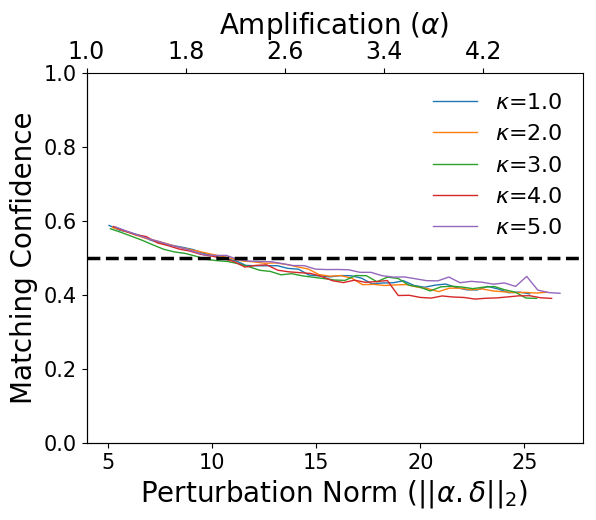}
		\caption{Face++}
		\vspace{0.1in}
		\label{fig:facepp_cw_rr}%
	\end{subfigure}
   \caption{In \emph{2018}: Transferability of uncropped images generated using CW attack}
    \label{fig:cw_rr}
\end{figure*}

\noindent{\bf CW:} Consistent with \S~\ref{sec:amp_margin}, transferability increases as amplification increases, and we observe transferability starting at $\norm{\alpha \cdot \delta}_2 \approx 6$ for Azure (Figure~\ref{fig:azure_cw_rr}), $\norm{\alpha \cdot \delta}_2 \approx 4.5$ for Rekognition (Figure~\ref{fig:awsverify_cw_rr}), and $\norm{\alpha \cdot \delta}_2 \approx 12$ for Face++ (Figure~\ref{fig:facepp_cw_rr}). The slope decreases beyond a specific point across all three models, suggesting that increasing the amplification factor $\alpha$ will only produce marginal privacy gains. We also observe limited correlation between the value of $\kappa$ and matching confidence. This observation suggests that the choice of $\alpha$ is more relevant for transferability than the choice of $\kappa$. Observe, however, that for a given value of $\alpha$, the transferability varies for different APIs.

\begin{figure*}[]
\centering
	\begin{subfigure}{0.31\textwidth}
		\centering
		\includegraphics[width=0.83\textwidth]{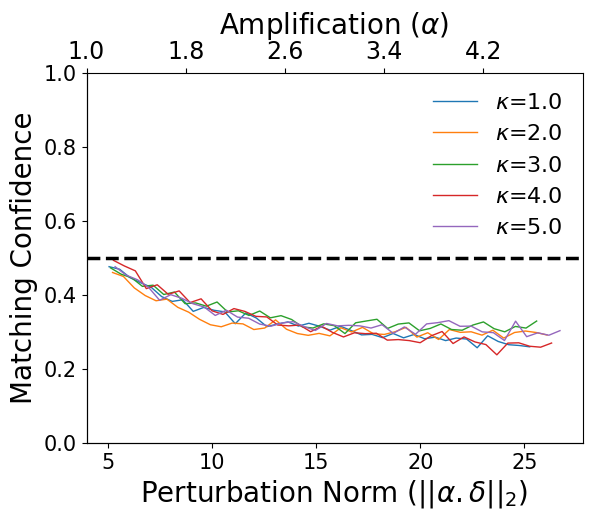}
        \caption{Azure}
        \vspace{0.1in}
		\label{fig:azure_cw_rr_2020}%
	\end{subfigure}
	\hfill
    \begin{subfigure}{0.31\textwidth}
		\centering
		\includegraphics[width=0.83\textwidth]{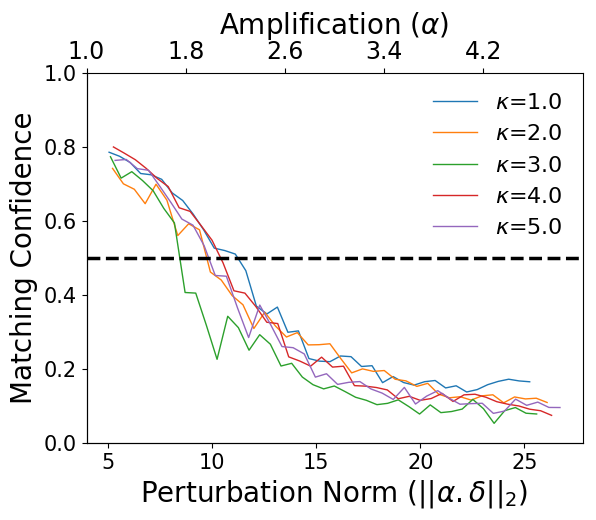}
		\caption{AWS Rekognition}
		\vspace{0.1in}
		\label{fig:awsverify_cw_rr_2020}%
	\end{subfigure}
	\hfill
    \begin{subfigure}{0.31\textwidth}
		\centering
		\includegraphics[width=0.83\textwidth]{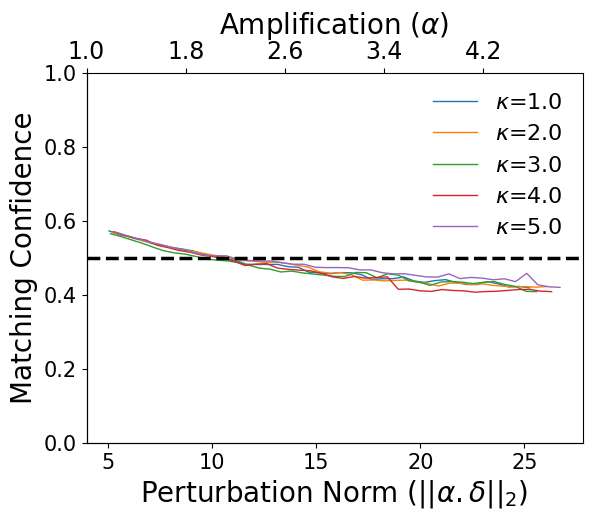}
		\caption{Face++}
		\vspace{0.1in}
		\label{fig:facepp_cw_rr_2020}%
	\end{subfigure}
    \caption{In \emph{2020}: Transferability of uncropped images generated using CW attack}
    \label{fig:cw_rr_2020}
\end{figure*}

\begin{figure*}[]
\centering
	\begin{subfigure}{0.31\textwidth}
		\centering
		\includegraphics[width=0.83\textwidth]{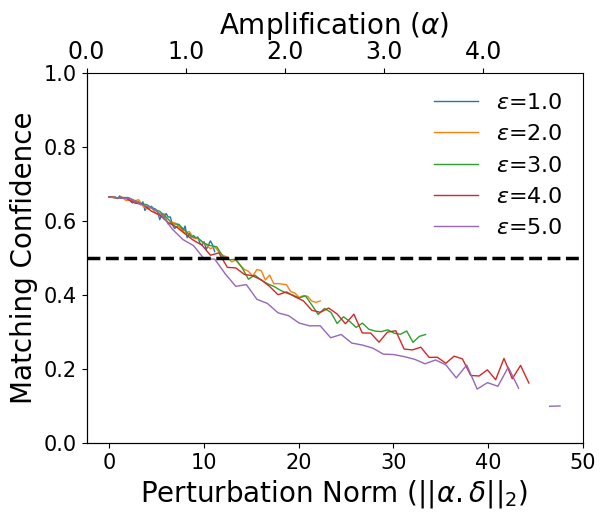}
        \caption{Azure}
        \vspace{0.1in}
		\label{fig:azure_pgd_rr}%
	\end{subfigure}
	\hfill
    \begin{subfigure}{0.31\textwidth}
		\centering
		\includegraphics[width=0.83\textwidth]{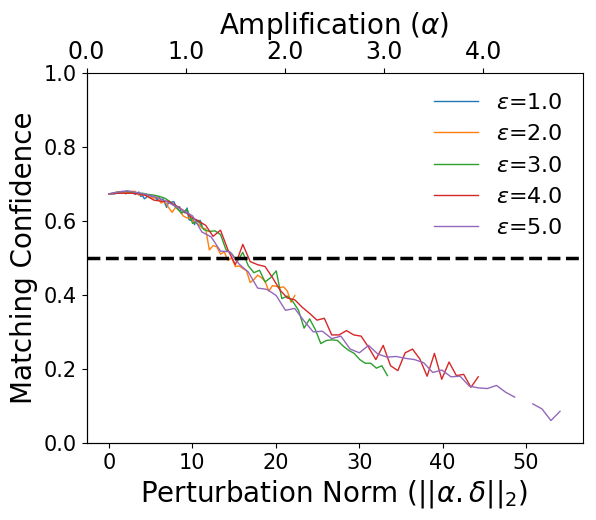}
		\caption{AWS Rekognition}
		\vspace{0.1in}
		\label{fig:awsverify_pgd_rr}%
	\end{subfigure}
	\hfill
    \begin{subfigure}{0.31\textwidth}
		\centering
		\includegraphics[width=0.83\textwidth]{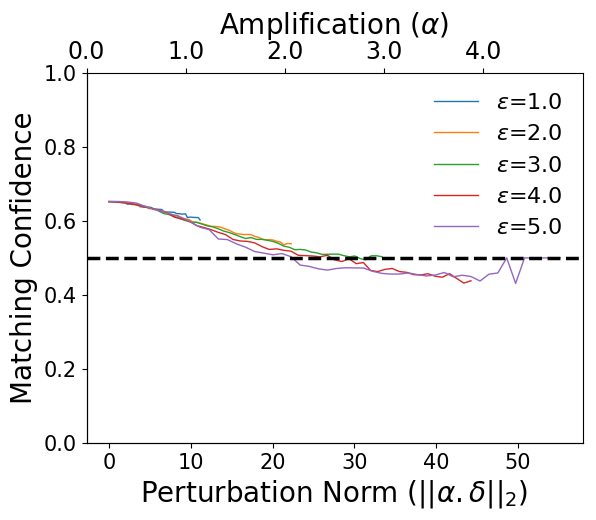}
		\caption{Face++}
		\vspace{0.1in}
		\label{fig:facepp_pgd_rr}%
	\end{subfigure}
  \caption{In \emph{2018}: Transferability of uncropped images generated using PGD attack}
    \label{fig:pgd_rr}
\end{figure*}

\vspace{1mm}
\noindent{\bf PGD:} Here, the choice of $\varepsilon$ has a greater impact on the rate of transferability (\ie how quickly, in terms of amplification, the confidence value reaches the $\tau=0.5$). The larger values of $\varepsilon$ correspond to faster transferability (steeper slope), as highlighted across all APIs in Figure~\ref{fig:pgd_rr}. As before, transferability across different models requires different levels of $\norm{\alpha \cdot \delta}_2$ (and consequently, $\alpha$). However, unlike images generated by the CW attack, the perturbation norm is larger for those generated by PGD.

\begin{figure*}[]
\centering
	\begin{subfigure}{0.3\textwidth}
		\centering
		\includegraphics[width=0.83\textwidth]{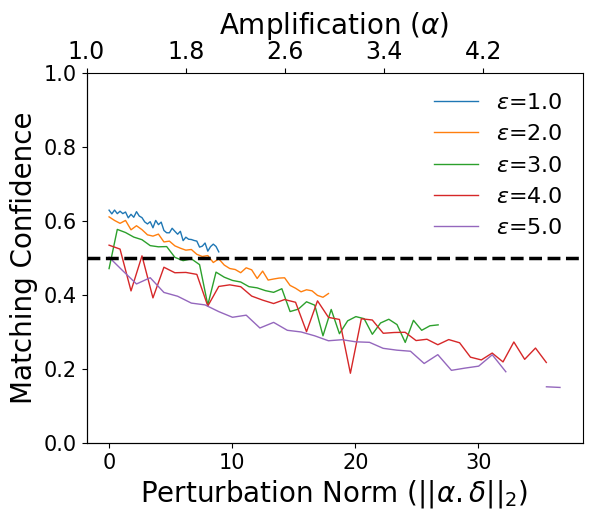}
        \caption{Azure}
        \vspace{0.1in}
		\label{fig:azure_pgd_rr_2020}%
	\end{subfigure}
	\hfill
    \begin{subfigure}{0.3\textwidth}
		\centering
		\includegraphics[width=0.83\textwidth]{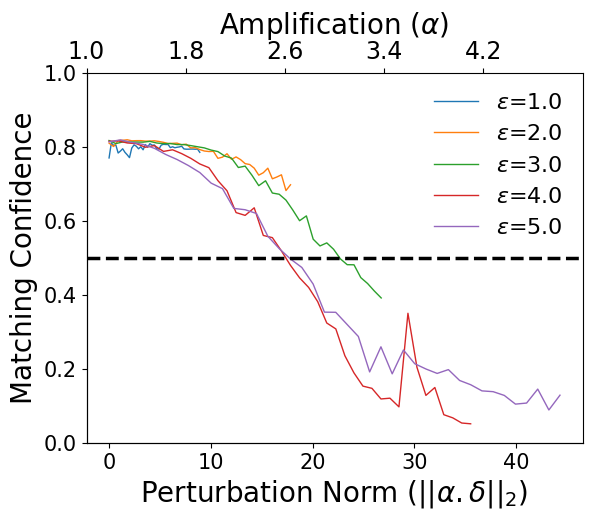}
		\caption{AWS Rekognition}
		\vspace{0.1in}
		\label{fig:awsverify_pgd_rr_2020}%
	\end{subfigure}
	\hfill
    \begin{subfigure}{0.3\textwidth}
		\centering
		\includegraphics[width=0.83\textwidth]{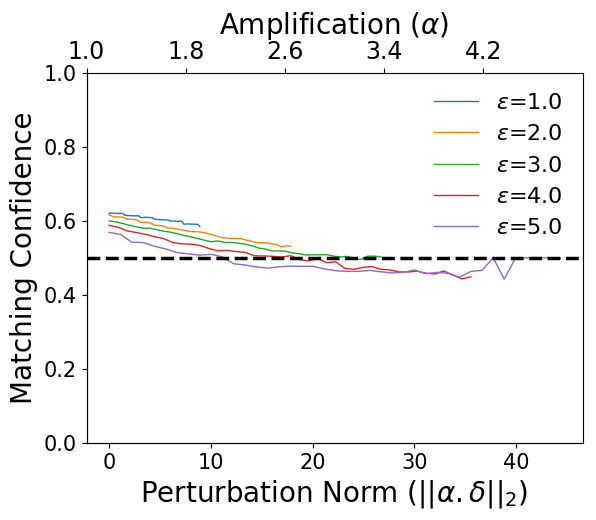}
		\caption{Face++}
		\vspace{0.1in}
		\label{fig:facepp_pgd_rr_2020}%
	\end{subfigure}
    \caption{In \emph{2020}: Transferability of uncropped images generated using PGD attack}
    \label{fig:pgd_rr_2020}
\end{figure*}

\subsection{Measuring API Robustness}
\label{subsec:robustness}

\noindent{\bf Longitudinal Study:} 
We conducted a longitudinal study to verify if commercial APIs have improved their robustness against adversarial samples. For this experiment, we utilize the images generated in August 2018 (from \S~\ref{subsec:commercial}) and verify if they transfer to the APIs in August 2020, {\em two years after initial testing}. We present the results in Figures~\ref{fig:cw_rr_2020} and Figures~\ref{fig:pgd_rr_2020} (for uncropped images modified using the exact same configuration of CW and PGD as in \S~\ref{subsec:commercial}). We observe that transferability persists across all 3 APIs (to variable degrees) despite the passage of time and potential retraining conducted by the API providers. The trends observed in \S~\ref{subsec:commercial} still hold. This fidning suggests that APIs {\em have not deployed mechanisms} to provide adversarial robustness.

\begin{table}[ht]
\begin{center}
  \begin{tabular}{c c c c c}
    \toprule
  \small{\bf Dataset} & \small{\bf Base ($\ell_2$)} & \small{\bf AT ($\ell_2$)} & \small{\bf Base} (cos) & \small{\bf AT} (cos)\\ 
    \midrule
    \midrule
    \small{LFW~\cite{huang2008labeled}} & 52.57\% & 38\% & 53.54\% & 38.47\% \\
    \small{VGGFace2~\cite{cao2018vggface2}} & 60.13\% & 48.22\% & 60.60\% & 48.25\% \\
     \small{Celeb~\cite{guo2016ms}} & 82.29\% & 80.41\% & 83.39\% & 81.03\% \\ 
    \bottomrule
    \end{tabular}
\end{center}
\caption{Top-1 accuracies after adversarial training. Note that adversarial training decreases top-1 accuracy. \textbf{Base} refers to the baseline natural accuracy (before adversarial training), and \textbf{AT} refers to the natural accuracy after adversarial training. $\ell_2$ denotes the 2-norm, and cos denotes the cosine similarity measure.}
\label{table:advtop}
\end{table}

\vspace{1mm}
\noindent{\bf Adversarial Training:} A curious reader may wonder if adversarial training~\cite{madry2017towards} can safeguard APIs against adversarial inputs, such as those generated through \name. It is well understood that, in classification settings, adversarial robustness is achieved at the expense of natural accuracy~\cite{tsipras2018robustness}. To empirically validate this observation in the context of metric embeddings, we train a variant of the FLVC model both naturally and adversarially for 3000 epochs on a subset of 50 labels (\ie 23435 images) sampled from the VGGFace2 dataset~\cite{cao2018vggface2}. We trained the model with a mixture of 50\% adversarial examples and 50\% natural examples. We generated adversarial examples using the $\ell_2$ variant of CW (and the hinge loss) with $\kappa=6$ and $\alpha=3$ on half of the labeled images in our training set. The robustness results are presented in Table~\ref{table:advtop}. The \textbf{Base} columns refer to the {\em natural accuracy} (\ie accuracy on clean/non-adversarial samples) when the model is trained naturally. The \textbf{AT} columns refer to the natural accuracy of the model when the model is trained adversarially. For natural accuracy, using three datasets for inference, we can observe that {\em in all cases}, the top-1 accuracy (measured using both 2-norm and cosine similarity) decreases with adversarial training. Thus, our findings are consistent with prior work in the classification regime~\cite{madry2017towards}. When we test the {\em adversarial accuracy} (\ie accuracy on adversarial samples) of the model using adversarial samples generated from VGGFace2, we observe an increase from 41.49\% to 48.59\% (when top-1 accuracy is calculated using the 2-norm) and from 41.70\% to 48.40\% (when top-1 accuracy is calculated using the cosine similarity measure)\footnote{These results are not presented in Table~\ref{table:advtop}.}. We report additional details and results in Appendix~\ref{app:atraining}.

\subsection{Top-n Recognition}
\label{sec:topn}

Finally, we study the efficacy of our targeted attack if the adversary uses top-n recognition instead of top-1 recognition. We generated masked inputs using (a) the FLVT model as the surrogate, (b) the hinge loss, and (c) the CW$_{large}$ attack for 6 of the 10 labels described earlier. We consider all pair-wise combinations for the following identities: \texttt{Matt, Leonardo, Barack, Morgan, Melania} and \texttt{Taylor}. Table~\ref{table:cw_topn} contains the results of top-3 recognition success for Face++, Azure, and AWS Rekognition (in sequence). Note that a success event is one where the correct label is in the top 3 labels predicted for the input. Each entry in the table is the average of the 30 pairings we consider. We observe that across all the APIs: (a) increasing $\kappa$ increases attack success (\ie, decreases accuracy), and (b) increasing $\alpha$ also increases attack success (\ie, decreases accuracy). Note that the trends we observe are consistent independent of the exact attack we use (as denoted by results for the $\ell_2$ variant of PGD using the FLVT model in Table~\ref{table:pgd_topn}), or the exact surrogate model--though they impact the magnitude of attack success.

\begin{table*}
\begin{tabular}{c}
\toprule
{} \\
{\bf $\kappa$} \\ 
\midrule
\midrule
{\bf 0} \\
{\bf 5} \\
{\bf 10} \\
\bottomrule
\end{tabular}
\hfill
\begin{tabular}{c c c c c}
\toprule
{} & {}& {\bf $\alpha$}& {}& {}\\
{\bf 1.8} & {\bf 3.4} & {\bf 5.0} & {\bf 6.6} & {\bf 8.2}\\    
\midrule
\midrule
    96\% & 78\% & 54\% & 40\% & 22\% \\
    100\%& 74\% & 58\% & 34\% & 24\% \\
    96\%& 74\% & 46\% & 32\% & 24\% \\
    \bottomrule
\end{tabular}
\hfill
\begin{tabular}{c c c c c}
\toprule
{} & {}& {\bf $\alpha$}& {}& {}\\
{\bf 1.8} & {\bf 3.4} & {\bf 5.0} & {\bf 6.6} & {\bf 8.2}\\    
\midrule
\midrule
    78\% & 34\% & 16\% & 12\% & 6\% \\
    80\%& 40\% & 20\% & 6\% & 4\% \\
    76\%& 30\% & 8\% & 6\% & 4\% \\
    \bottomrule
\end{tabular}
\hfill
\begin{tabular}{c c c c c}
\toprule
{} & {}& {\bf $\alpha$}& {}& {}\\
{\bf 1.8} & {\bf 3.4} & {\bf 5.0} & {\bf 6.6} & {\bf 8.2}\\    
\midrule
\midrule
    95\% & 75\% & 55\% & 40\% & 35\% \\
    95\%& 75\% & 50\% & 35\% & 35\% \\
    95\%& 65\% & 45\% & 35\% & 35\% \\
    \bottomrule
\end{tabular}
\caption{Top-3 recognition accuracy (using the 2-norm) for Face++, Azure, AWS Rekognition respectively. Masked samples were generated using the FLVT model as the surrogate and the CW$_{large}$ attack (refer Table~\ref{table:params}).}
\label{table:cw_topn}
\end{table*}

\begin{table*}
\begin{tabular}{c}
\toprule
{} \\
{\bf $\kappa$} \\ 
\midrule
\midrule
{\bf 0} \\
{\bf 5} \\
{\bf 10} \\
\bottomrule
\end{tabular}
\hfill
\begin{tabular}{c c c c c}
\toprule
{} & {}& {\bf $\alpha$}& {}& {}\\
{\bf 1.8} & {\bf 3.4} & {\bf 5.0} & {\bf 6.6} & {\bf 8.2}\\    
\midrule
\midrule
    95\% & 70\% & 15\% & 0\% & 0\% \\
    95\%& 65\% & 20\% & 0\% & 0\% \\
    80\%& 55\% & 0\% & 0\% & 0\% \\
    \bottomrule
\end{tabular}
\hfill
\begin{tabular}{c c c c c}
\toprule
{} & {}& {\bf $\alpha$}& {}& {}\\
{\bf 1.8} & {\bf 3.4} & {\bf 5.0} & {\bf 6.6} & {\bf 8.2}\\    
\midrule
\midrule
    75\% & 20\% & 5\% & 0\% & 0\% \\
    75\%& 15\% & 5\% & 0\% & 0\% \\
    70\%& 10\% & 0\% & 0\% & 0\% \\
    \bottomrule
\end{tabular}
\hfill
\begin{tabular}{c c c c c}
\toprule
{} & {}& {\bf $\alpha$}& {}& {}\\
{\bf 1.8} & {\bf 3.4} & {\bf 5.0} & {\bf 6.6} & {\bf 8.2}\\    
\midrule
\midrule
    90\% & 75\% & 60\% & 40\% & 10\% \\
    85\%& 75\% & 70\% & 20\% & 0\% \\
    85\%& 75\% & 60\% & 40\% & 10\% \\
    \bottomrule
\end{tabular}
\caption{Top-3 recognition accuracy (using the 2-norm) for Face++, Azure, AWS Rekognition respectively. Masked samples were generated using the FLVT model as the surrogate and the PGD attack (refer Table~\ref{table:params}).}
\label{table:pgd_topn}
\end{table*}

\section{User Study}
\label{sec:user}

Thus far, we have studied the efficacy of our approach on transferability. In this section, we check if the perturbed (and amplified) images are {\em user-friendly}, \ie if the users are willing to upload such images to social media platforms. In the first study (\S~\ref{sec:prefixed}), users see images amplified by different values of $\alpha$ and are asked if they would upload said images. In the second study (\S~\ref{sec:user_upload}), users upload images of their choice to \name's online service. After the service returns a perturbed image, we ask the user to ascertain the utility of this image. Both studies are approved by our IRB and are conducted on Amazon's MTurk platform. The main difference between the two studies is the control condition. The first study controls the images to assess the user's perception of the perturbation at the potential cost of the results' ecological validity. The second study allows users to upload images of their choice to get more realistic assessments.

\subsection{Perturbation Tolerance}
\label{sec:prefixed}

\begin{figure*}[t]%
\centering
    \begin{subfigure}{0.13\textwidth}
        \centering
        \includegraphics[width=\textwidth]{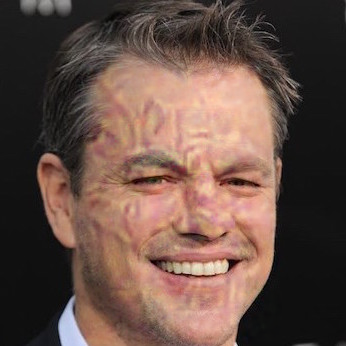}
        \caption{$\alpha=1.6$}
        \vspace{0.1in}
        \label{fig:portrait_amp_0.0}%
    \end{subfigure}
    \begin{subfigure}{0.13\textwidth}
        \centering
        \includegraphics[width=\textwidth]{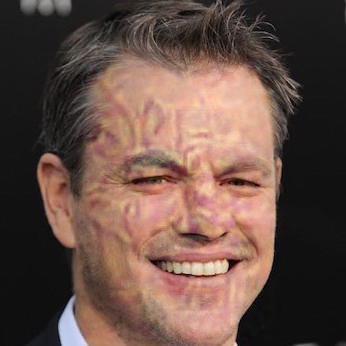}
        \caption{$\alpha=$1.7}
        \vspace{0.1in}
        \label{fig:portrait_amp_1.0}%
    \end{subfigure}
    \begin{subfigure}{0.13\textwidth}
        \centering
        \includegraphics[width=\textwidth]{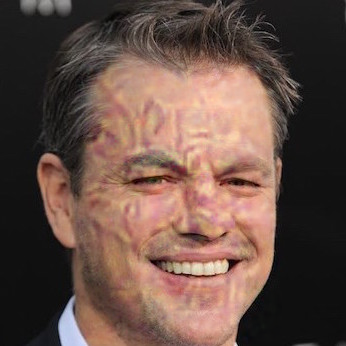}
        \caption{$\alpha=$1.8}
        \vspace{0.1in}
        \label{fig:portrait_amp_1.2}%
    \end{subfigure}
    \begin{subfigure}{0.13\textwidth}
        \centering
        \includegraphics[width=\textwidth]{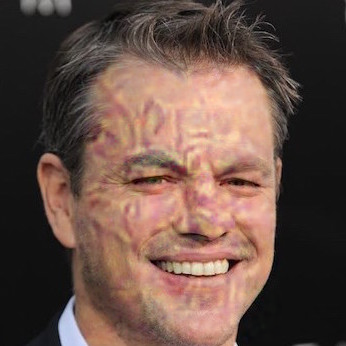}
        \caption{$\alpha=$1.9}
        \vspace{0.1in}
        \label{fig:portrait_amp_1.4}%
    \end{subfigure}
    \begin{subfigure}{0.13\textwidth}
        \centering
        \includegraphics[width=\textwidth]{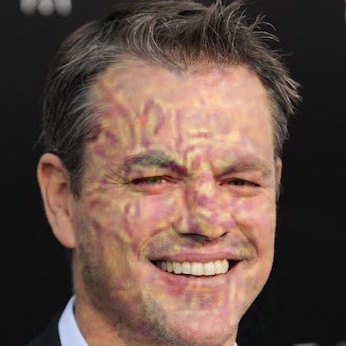}
        \caption{$\alpha=$2.0}
        \vspace{0.1in}
        \label{fig:portrait_amp_1.6}%
    \end{subfigure}
    \begin{subfigure}{0.13\textwidth}
        \centering
        \includegraphics[width=\textwidth]{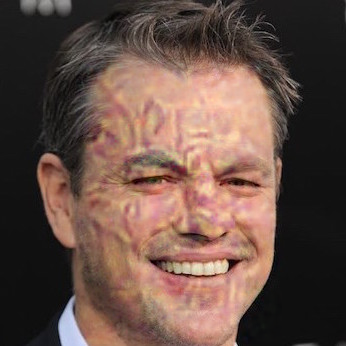}
        \caption{$\alpha=$2.1}
        \vspace{0.1in}
        \label{fig:portrait_amp_1.8}%
    \end{subfigure}
    \begin{subfigure}{0.13\textwidth}
        \centering
        \includegraphics[width=\textwidth]{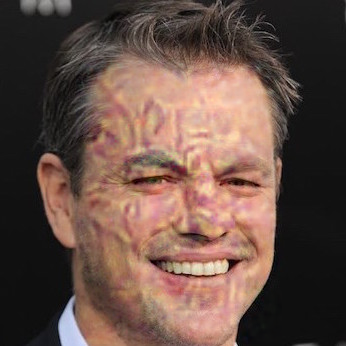}
        \caption{$\alpha=$2.2}
        \vspace{0.1in}
        \label{fig:portrait_amp_2.0}%
    \end{subfigure}
    \caption{Photos used in the user study for portrait case with the perturbation increasingly amplified.}
    \label{fig:portrait}
\end{figure*}

\begin{figure*}[t]%
\centering
    \begin{subfigure}{0.13\textwidth}
        \centering
        \includegraphics[width=\textwidth]{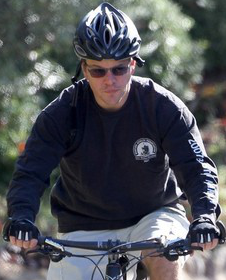}
        \caption{Original}
        \vspace{0.1in}
        \label{fig:bike_amp_0.0}%
    \end{subfigure}
    \begin{subfigure}{0.13\textwidth}
        \centering
        \includegraphics[width=\textwidth]{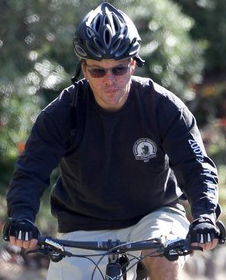}
        \caption{$\alpha=1$}
        \vspace{0.1in}
        \label{fig:bike_amp_1.0}%
    \end{subfigure}
    \begin{subfigure}{0.13\textwidth}
        \centering
        \includegraphics[width=\textwidth]{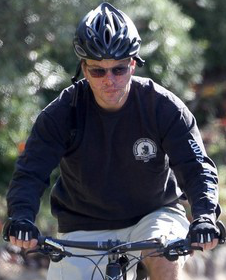}
        \caption{$\alpha=1.2$}
        \vspace{0.1in}
        \label{fig:bike_amp_1.2}%
    \end{subfigure}
    \begin{subfigure}{0.13\textwidth}
        \centering
        \includegraphics[width=\textwidth]{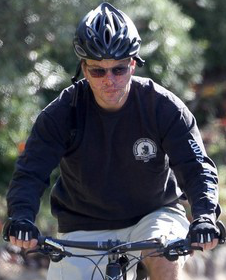}
        \caption{$\alpha=1.4$}
        \vspace{0.1in}
        \label{fig:bike_amp_1.4}%
    \end{subfigure}
    \begin{subfigure}{0.13\textwidth}
        \centering
        \includegraphics[width=\textwidth]{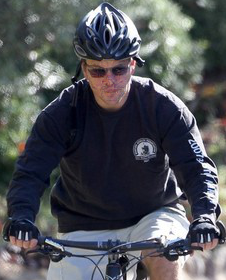}
        \caption{$\alpha=1.6$}
        \vspace{0.1in}
        \label{fig:bike_amp_1.6}%
    \end{subfigure}
    \begin{subfigure}{0.13\textwidth}
        \centering
        \includegraphics[width=\textwidth]{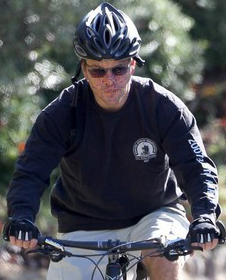}
        \caption{$\alpha=1.8$}
        \vspace{0.1in}
        \label{fig:bike_amp_1.8}%
    \end{subfigure}
    \begin{subfigure}{0.13\textwidth}
        \centering
        \includegraphics[width=\textwidth]{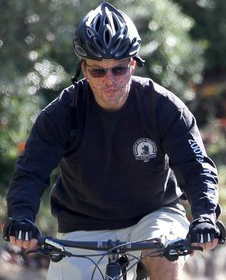}
        \caption{$\alpha=2$}
        \vspace{0.1in}
        \label{fig:bike_amp_2.0}%
    \end{subfigure}
    \caption{Photos used in the user study for the background case with the perturbation increasingly amplified. The perturbation is focused on the face region.}
    \label{fig:bike}
\end{figure*}

We conducted the first user study to assess \textit{user-perceived utility} of the images that \name generates. We performed this assessment for samples generated using $\ell_2$ variants of both CW and PGD attacks (as described in \S~\ref{subsec:commercial}). Through this study, we aim to understand the amount of perturbation users are willing to tolerate. We consider two types of images: (a) {\em portrait}, where the face is the focus of the image (Figure~\ref{fig:portrait}), and (b) {\em background}, where the background is the focus of the image, and the face is in the image (Figure~\ref{fig:bike}). We summarize our findings below:

\begin{itemize}
\itemsep0em
\item Privacy-conscious individuals are willing to tolerate larger perturbation levels for improved privacy.
\item For the portrait image, 40\% of respondents had no problem uploading a perturbed image to the social media platform (the exact tolerable perturbation level, however, differed among respondents).
\item For the background image, the vast majority of users exhibited tolerance to higher image perturbation.
\end{itemize}

\subsubsection{Study Design}

\noindent{\textbf{Participant Recruitment}:} We recruited a total of 167 and 163 Amazon MTurk {\em master} workers for the portrait and background study, respectively. With this number of users, each image in the portrait study received at least five ratings, and each image in the background study received 163 ratings. Each worker was compensated \$1 for their effort, with an average completion time of 6 minutes. We present the demographics of the participants in Table~\ref{table:demos} in Appendix~\ref{app:user}.

\vspace{1mm}
\noindent{\textbf{Study Protocol}:} We asked the user to rate a different image every time. For each participant in the portrait group, we display (a) 20 random images (each on a different page) where $\alpha \in [1.4,2.4]$. For participants in the background group, we show the same 20 images where $\alpha \in [1.4,2.4]$. We choose this range as it enables (some degree of) transferability (as witnessed in Figures~\ref{fig:cw_rr} and~\ref{fig:cw_rr_2020}).

After rating the images, we asked the respondents a set of four questions on a 5-point Likert scale to gauge their privacy concern levels. We utilize the ``Concern for Privacy"~\cite{privacy_scale:2004} scale which is modeled after the well-known ``Information Privacy Concern scale" of Smith \etal~\cite{privacy_scale:1996}. We use this set of questions to divide the respondents into two groups: (a) \textit{Privacy Conscious (PC)}, and (b) \textit{Not Privacy Conscious (NPC)}. Respondents belonging to the first group are those who have answered all four questions with either \texttt{Strongly Agree} (SA) or \texttt{Agree} (A). The second group of respondents is those who responded with \texttt{Neutral} (N), \texttt{Disagree} (D), or \texttt{Strongly Disagree} (SD) any of the questions. Finally, we require the respondents to answer a set of general demographic questions. The respondents were made aware that these questions are optional and require no personally identifiable information. 

\vspace{1mm}
\noindent{\bf Image Evaluation:} For each displayed image, we asked the respondents to answer the following questions: (a) ``\textit{I have no problem in uploading this photo to social media}", and (b) ``\textit{I would upload the image to social media to prevent automatic tagging of my face}" on a 5-point Likert scale. %

\begin{figure}[t]%
\centering
    \begin{subfigure}{0.24\textwidth}
        \centering
        \includegraphics[width=\textwidth]{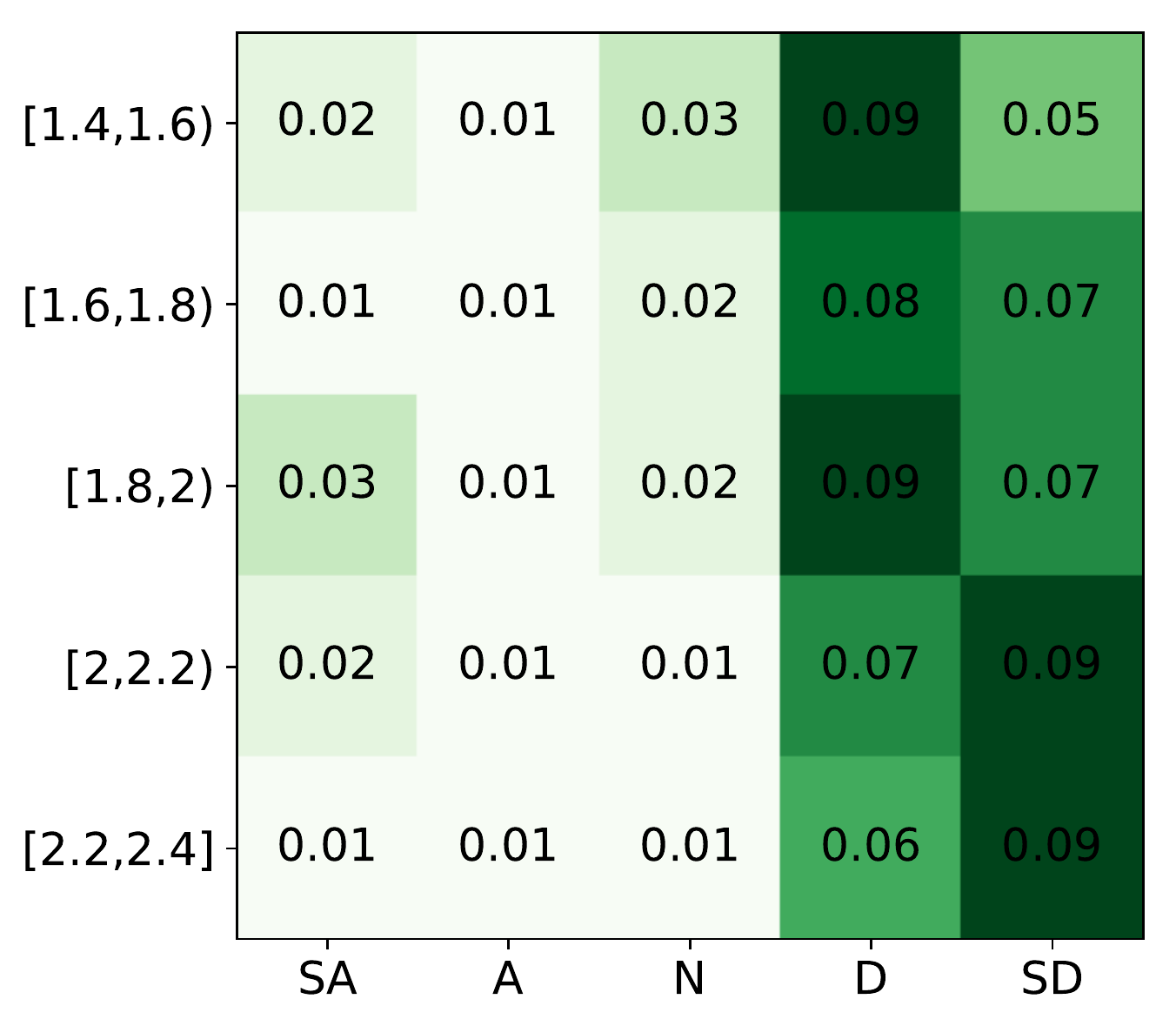}
        \caption{Not Privacy Conscious}
        \vspace{0.1in}
        \label{fig:npc_p}%
    \end{subfigure}
\begin{subfigure}{0.24\textwidth}
        \centering
        \includegraphics[width=\textwidth]{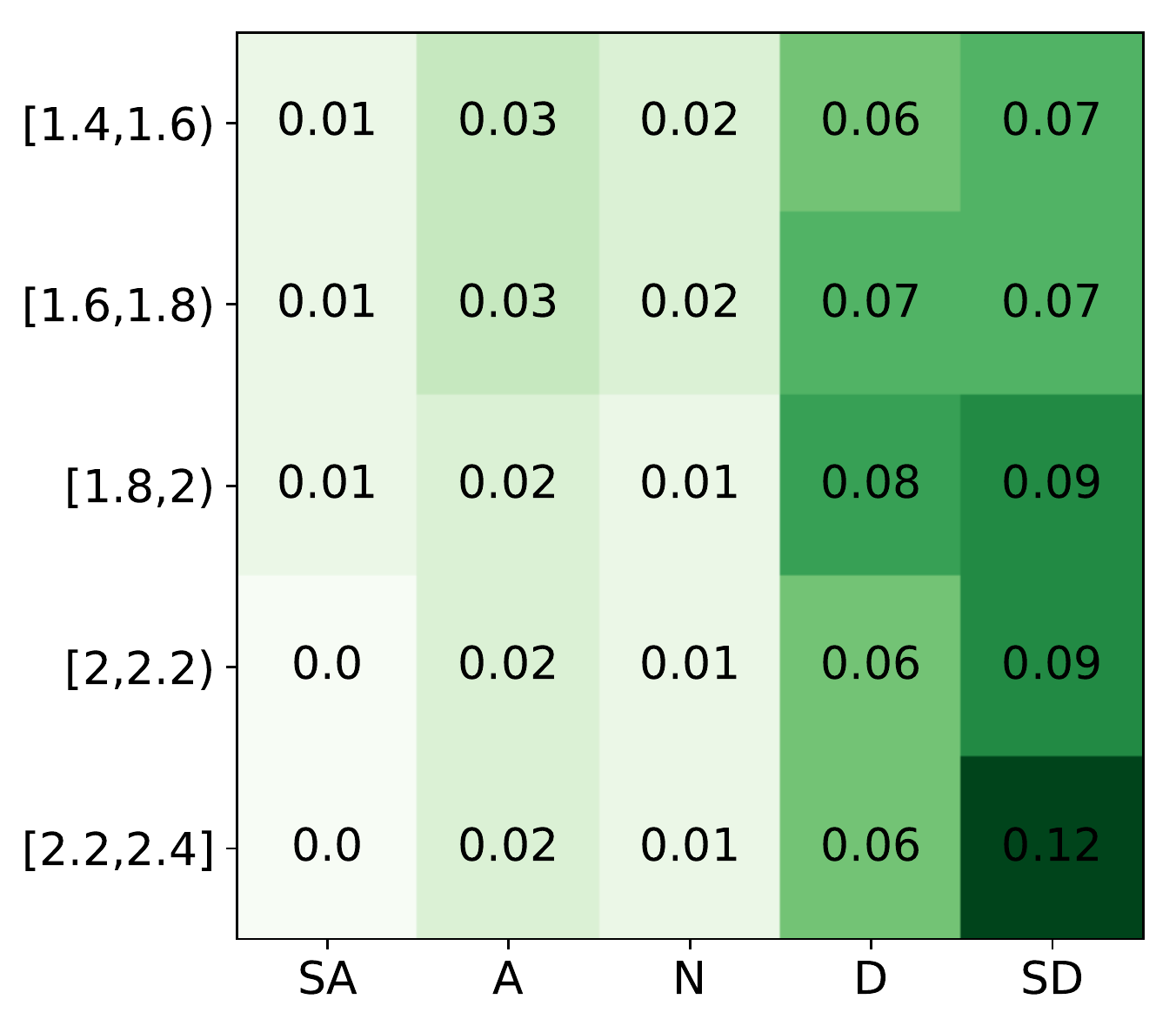}
        \caption{Privacy Conscious}
        \vspace{0.1in}
        \label{fig:pc_p}%
    \end{subfigure}
    \caption{The distribution of responses for the \textbf{portrait} scenario for the PC and NPC groups. Each cell value contains the portion of responses for a specific $\alpha$ range and user satisfaction value.}
    \label{fig:portrait_responses}
\end{figure}

\subsubsection{Results}

\noindent{\bf 1. Portrait Images:} Each user was shown a set of images with a corresponding $\alpha$ value unknown to the user. We grouped user responses into 5 buckets corresponding to the ranges $[1.4,1.6), [1.6,1.8), [1.8,2),[2,2.2),[2.2,2.4]$. These five buckets represent increasing levels of perturbation. 

We first discuss results for the NPC category (Figure~\ref{fig:npc_p}); as $\alpha$ increases, the number of users who do not wish to upload the image (the SD category \ie the last column) also increases. The inverse is also true; as the perturbation is lower, the number of participants who wish to upload the image is higher (\ie the first column). Similar observations can be made for the PC category (Figure~\ref{fig:pc_p}). Combining both groups, we found that 40\% of the respondents are impartial to uploading at least one of the perturbed images. It is worth noting that portrait images are a unique case, where the subject is the highlight of these images, with a minimal background context (Figure~\ref{fig:portrait_responses}). Thus, the perturbation is more explicitly visible on the user's face, accounting for the received responses. 

Nevertheless, we observe that the PC respondents are slightly more inclined to accept perturbed images than NPC users. This observation is evident from comparing Figure~\ref{fig:npc_p} and Figure~\ref{fig:pc_p}, where the latter figure shows a higher density of responses in the SA-A region. Further, we find the user perception to be dependant on privacy consciousness in all five amplification ranges ($p<0.05$ according to the $\chi^2$ test after applying the Holm-Bonferroni~\cite{holm-comp} method to correct for multiple comparisons).

\begin{figure}[t]%
\centering
    \begin{subfigure}{0.24\textwidth}
        \centering
        \includegraphics[width=\textwidth]{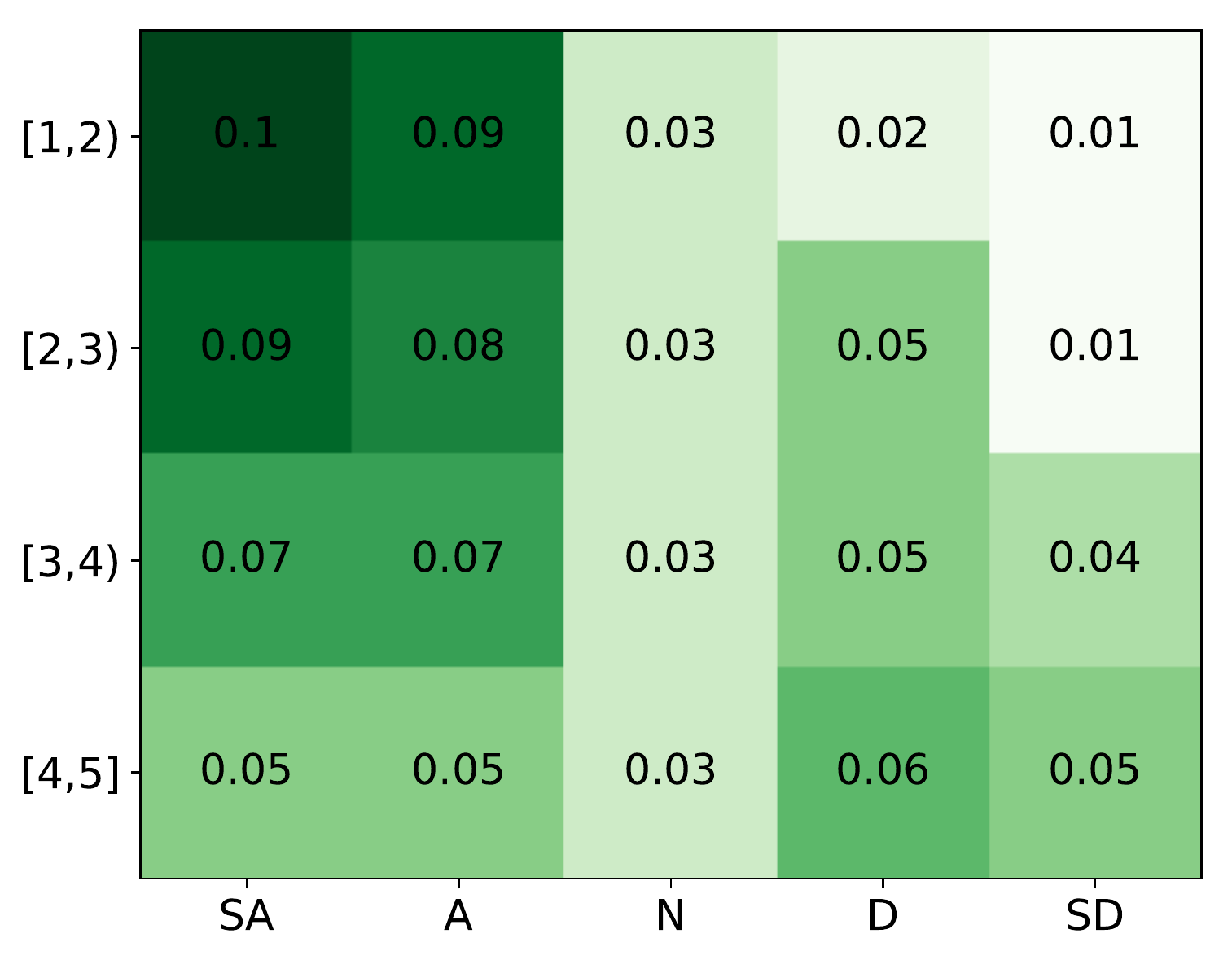}
        \caption{Not Privacy Conscious}
        \vspace{0.1in}
        \label{fig:npc_b}%
    \end{subfigure}
\begin{subfigure}{0.24\textwidth}
        \centering
        \includegraphics[width=\textwidth]{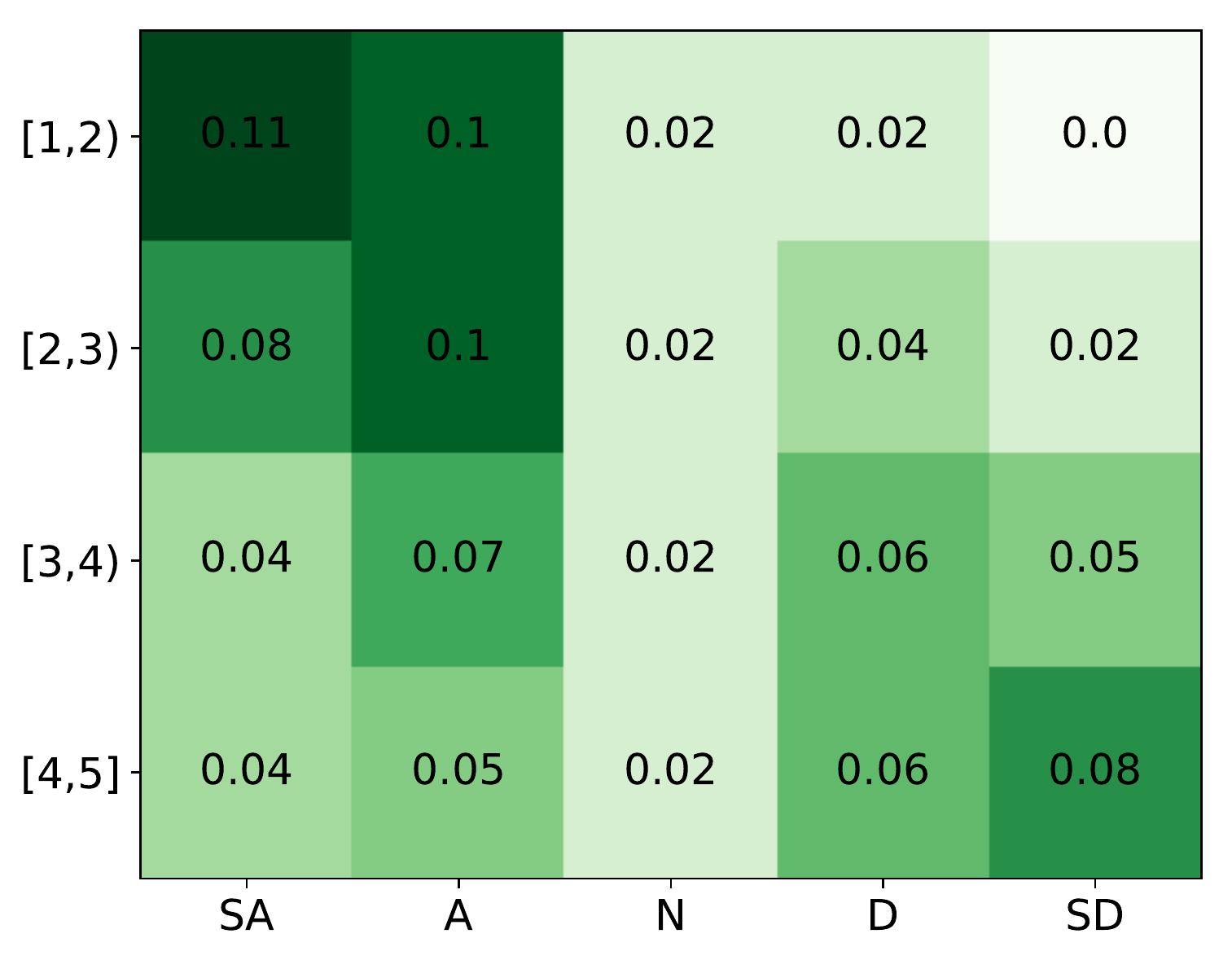}
        \caption{Privacy Conscious}
        \vspace{0.1in}
        \label{fig:pc_b}%
    \end{subfigure}
    \caption{The distribution of responses for the \textbf{background} scenario for the PC and NPC groups. Each cell value contains the portion of responses for a specific $\alpha$ range and user satisfaction value.}
    \label{fig:bike_responses}
\end{figure}

\vspace{1mm}
\noindent{\bf 2. Background Images:} Figure~\ref{fig:bike_responses} shows the user responses for background images to be far more favorable than the portrait case. As evident from Figure~\ref{fig:bike}, the face constitutes a small region of such images, with other relevant features. Thus, resizing, amplifying, and adding the adversarial perturbation does not make as noticeable a difference as in the portrait image case. Except for the last range of $\alpha$ values, most of the respondents agreed to upload the perturbed image to the social media platform. In the first three ranges of $\alpha$, we did not observe privacy consciousness to be a factor in the respondents' answers ($p>0.05$ according to the $\chi^2$ test after applying the Holm-Bonferroni method to correct for multiple comparisons). In one case, one of the respondents indicated that they do not observe any difference in the images and wondered whether we were testing respondents by showing the same image for every question. The only exception was the last $\alpha$ range, where large image perturbation is high enough to be unacceptable to our PC respondents. 

Finally, the choice of the image exhibits a clear distinction in the user's perception of the perturbation ($p=0$ according to the $\chi^2$ test when comparing background and portrait responses over the same ranges of $\alpha$). This distinction holds for all respondents. Users are typically more interested in preserving their privacy in situations related to a certain activity, behavior, or social context~\cite{chi:2010}. A portrait image contains little context about user activity or behavior. On the other hand, background images contain more context related to user activity, behavior, location, and social circles. For these images, users have a high incentive to avoid being automatically tagged and tracked by social media platforms.

In summary, we observed that \name helps balance the privacy-utility trade-off of users. Most of the respondents have no problem uploading the background image, regardless of their privacy stance. Even for portrait images, a part of PC respondents accepted uploading perturbed images. 

\subsection{End-to-End Usability}
\label{sec:user_upload}

While the first study suggests that privacy-conscious users are willing to upload perturbed images, the images themselves were not relevant to the users. We address this shortcoming through another user study with a more realistic setting, where users upload images to \name's online portal. This study design improves the ecological validity from the first study as we show the users perturbed versions of images relevant to them. We describe the specifics of the study below.

\subsubsection{Study Design}
\vspace{2mm}

\noindent{\bf Participant Recruitment:} We recruited a total of 100 Amazon MTurk workers. Each worker was compensated \$2 for their effort, with an average completion time of 10 minutes. After filtering responses that fail our attention checkers, we report the results based on 93 participants. We present the demographics of the participants in Table~\ref{table:demos} in Appendix~\ref{app:user}.

\vspace{1mm}
\noindent{\bf Study Protocol:} We used a between-subject study by asking each user to first upload an image of their choice to our portal (Appendix~\ref{app:screenshots}). The only requirements were to ensure that each image contained a person/people of significance, and the faces of these people were easily identifiable. The uploaded face is compared with a similar target identity (in the embedding space), which is used for the attack. The service randomly assigns the user a value of $\alpha$\footnote{$\kappa$ is fixed to be 10.}, and proceeds to return a perturbed variant of the user-uploaded image. We avoided explicit mentions of privacy in the survey's introduction to reduce priming effects. Then, the user answers the same questions as in \S~\ref{sec:prefixed}--to understand if they are tolerant of the perturbation and willing to upload the masked input, and to ascertain their privacy preference. Similar to the previous study, participants were grouped into two categories: {\em Privacy Conscious (PC)} or {\em Not Privacy Conscious (NPC)}.

\subsubsection{Results}

\begin{figure}[t]%
\centering
    \begin{subfigure}{0.22\textwidth}
        \centering
        \includegraphics[width=\textwidth]{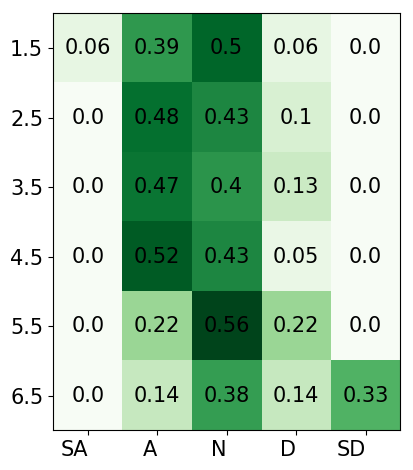}
        \caption{Not Privacy Conscious}
        \vspace{0.1in}
        \label{fig:uu_npc}%
    \end{subfigure}
\begin{subfigure}{0.22\textwidth}
        \centering
        \includegraphics[width=\textwidth]{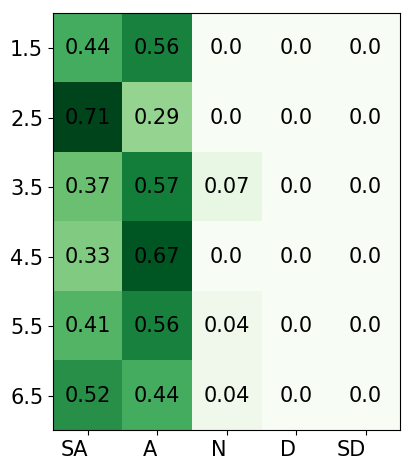}
        \caption{Privacy Conscious}
        \vspace{0.1in}
        \label{fig:uu_pc}%
    \end{subfigure}
    \caption{The distribution of responses for the scenario where the users upload images to be be perturbed. Each cell value contains the portion of responses for a specific $\alpha$ value and user satisfaction value.}
    \label{fig:user_upload}
\end{figure}

 The distribution of individual responses from this study can be found in Figure~\ref{fig:uu_npc} for PC participants, and Figure~\ref{fig:uu_pc} for NPC participants. For the PC group, we found that the value of $\alpha$ has little effect on the user's decision to upload the image. For all the values of $\alpha$, nearly all the users in the PC group agree to upload the perturbed image to a social media platform. For the NPC, we observe a shift in the users' decision; these users are likely to disagree to upload the perturbed photos for larger values of $\alpha$. Still, for this group, most participants fall within the neutral and agree categories, indicating the acceptability of the perturbations.  

We observe that the user responses in the second study followed a similar trend to those in the background scenario (Figure~\ref{fig:bike_responses} from \S~\ref{sec:prefixed}). While we did not have access to the uploaded photos for privacy reasons\footnote{Photos were immediately deleted after being processed by our online service.}, we conjecture that the users uploaded images feature other objects as well as faces, similar to the background case of \S~\ref{sec:prefixed}.

\section{Discussion}
\label{sec:discussion}

Here, we state some observations and limitations of \name. We stress that the findings we describe below are not conclusive and simply mirror our experiences with experimentation with various datasets, compression strategies, facial recognition models, and online APIs. 

\subsection{Observations}
\label{subsec:compression}

First, we observed that gender and race appear to play an important role in determining the target label should one use the targeted attack. The embeddings of people belonging to the same gender or race are closer in the embedding space (across all surrogate models we use).
Also, despite extensively perturbing particular identities (using large values of $\alpha$), such as those belonging to minorities, these perturbations do not transfer to the cloud APIs. This observation may suggest bias in the training data used by both victim and surrogate models, and has been thoroughly investigated in prior work~\cite{buolamwini2018gender}.

\begin{figure}[ht]%
\centering
    \includegraphics[width=0.75\columnwidth]{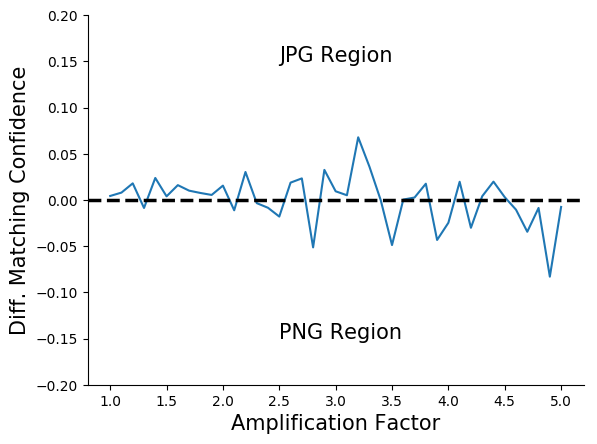}
    \caption{Masked samples survive compression.}
    \label{fig:jpg_v_png}
\end{figure}

Second, we observed that our masked samples survive lossy compression at the expense of modest losses in privacy (reduction of matching confidence). In Figure~\ref{fig:jpg_v_png}, we plot the difference in confidence values (on the Azure Face API) between a test image and the generated masked sample when the masked sample is passed as a PNG (lossless compression) and a JPG (lossy compression). The upper half of the plot are regions where the JPG has lower confidence than the PNG image, and vice-versa. We observe that the magnitude of the difference in confidence values is minimal, suggesting that the choice of compression standard does not impact the results.

\subsection{Limitations}
\label{subsec:limitations}

We highlight some limitations of \name, which we hope to address in future work. 

\begin{enumerate}
\itemsep0em
\item Like other black-box attack schemes~\cite{tramer2017spaces,liu2016delving,szegedy2013intriguing,papernot2016transferability,papernot2016practical}, our approach does not provide guarantees on transferability. Even with high values of $\alpha$, the masked samples may still not transfer. However, our approach always enhances privacy if privacy were to be measured by the decreasing confidence with which these metric networks are able to match faces.  
In addition, amplification offers users the flexibility to balance the trade-off between privacy and utility; PC users can increase amplification and obtain greater privacy at the expense of more visible perturbation.
\item While the time required for generating the masked sample can potentially bottleneck real-time image upload (refer Appendix~\ref{app:runtime}), we envision alternate deployment strategies, such as offline masked sample generation, to circumvent this bottleneck. 
\item A scenario that \name cannot circumvent is when other people on social media platforms tag faces. This provides the social media platform an additional signal (and some feedback) to fix incorrect predictions.
\item The most significant limitation of our work, similar to all other adversarial example generation strategies, is the ever-improving robustness of black-box models~\cite{cohen2019certified}. However, as we show in \S~\ref{subsec:robustness}, increasing robustness is at the expense of natural accuracy.
\item In practice, determining the right surrogate model to use to maximize transferability is a challenging problem. Right now, we exhaustively try all candidates. The same can be said for choosing the {\em optimal} target label.
\end{enumerate}

\section{Related Work}
\label{sec:related}

We discuss relevant work below. These can broadly be classified as work related to generating adversarial examples in a black-box setting, and work designed to preserve privacy in online platforms.

\subsection{Black-Box Attacks}

Several prior works demonstrate that some adversarial examples generated for one model may also be misclassified by another model~\cite{szegedy2013intriguing,papernot2016transferability, papernot2016practical}. For example, Papernot \etal~\cite{papernot2016practical}, propose an attack strategy of training a local model (as a surrogate for the black-box victim) using synthetically generated inputs. The victim DNN labels these inputs. The adversarial examples generated by the surrogate are shown to be likely misclassified by the target DNN as well.

Another line of work does not utilize the black-box models for the example generation process \ie the black-box model is never queried; work from Moosavi-Dezfooli \etal showed the existence of a universal perturbation for each model which can transfer across different images~\cite{moosavi2017universal}. Tramer \etal conducted a study of the transferability of different adversarial instance generation strategies applied to different models~\cite{tramer2017spaces}. The authors also proposed to use an ensemble of multiple models to generate adversarial examples to obtain increased transferability~\cite{tramer2017ensemble}. In a similar vein, Rajabi \etal~\cite{rajabi2021impractical} propose an approach to generate a universal perturbation (generated in a black-box manner) to be applied to all images. Finally, Sabour \etal~\cite{sabour2015adversarial} propose an approach to generate embedding perturbations, but in the white-box setting.

\subsection{Privacy}

Prior to our work, initial explorations have been made to utilize adversarial examples for protecting visual privacy~\cite{jia2019defending}. Raval \etal developed a perturbation mechanism that jointly optimizes privacy and utility objectives~\cite{raval2017protecting}. Targeting face recognition systems, Sharif \etal developed a physical attack approach~\cite{glass1, glass2}. The proposed algorithm first performs an adversarial attack on digital face images and limits the perturbation to an eyeglass frame-shaped area. Then the adversarial perturbation is printed into a pair of physical eyeglasses and can be worn by a person to dodge face detection, or to impersonate others in these face recognition systems. Being able to bring an adversarial attack into the physical world, this approach preserves visual privacy against face recognition. Additional prior work~\cite{sun2018natural,sun2018hybrid} operate in the classification setting, where the loss objective (and attack formulation) are different from ours. Work by McPherson \etal~\cite{mcpherson2016defeating} is a solution to an orthogonal problem, where the perturbations added are structured and human perceptible. 

The work of Bose \etal \cite{bose2018adversarial} attempts to induce failure events given black-box access to {\em facial detectors}. Given white-box access to a face detector, the proposed scheme trains a generator against it for a given image. The generated adversarial perturbation aims to dodge the face detector so that the faces are not detected. Concurrent work from Shan \etal~\cite{shan2020fawkes} explores the same problem. Using data poisoning attacks, they obfuscate faces at high success rates. In doing so, their approach could have an impact on large face recognition models trained using public images. However, their threat model differs in that they rely on online services using user data {\em to train} their deep learning models. However, social media platforms may opt to use a pre-trained model for face recognition tasks to avoid retraining on potentially malicious images. \name operates at test time, so it transfers {\em regardless of the platform's training policy}.

\section{Conclusion}

In this paper, we present \name, a system designed to preserve user privacy from facial recognition services. We design two new loss functions to attack metric learning systems, and extensively evaluate our approach using various models, architectures, parameters, and hyper-parameters across three commercial face recognition APIs. Our results affirm the utility of \name. Through our evaluation, we observe several artifacts that suggest training dataset, and algorithmic bias against specific sub-populations.

\section{Acknowledgements}
The authors would like to thank the reviewers for their constructive reviews that have considerably improved this work.
Suman Banerjee, Varun Chandrasekaran, and Chuhan Gao are/were supported in part through the following US NSF grants: CNS-1838733, CNS-1719336, CNS-1647152, CNS-1629833, and CNS-2003129. Kassem Fawaz and Somesh Jha are supported by DARPA under agreement number 885000. 

\newpage
\bibliographystyle{IEEEtran}
\bibliography{references}

\newpage
\newpage
\onecolumn
\appendix
\subsection{Demographics}
\label{app:user}

We report the demographic information of the participants of our user studies (refer \S~\ref{sec:user}) in Table~\ref{table:demos}. Columns \textbf{P} and \textbf{B} refer to Portrait and Background studies (refer \S~\ref{sec:prefixed}) and column \textbf{UU} refers to the User Uploaded images study (refer \S~\ref{sec:user_upload}). 

\begin{table}[H]
\begin{center}
  \begin{tabular}{  l  c  c  c}
    \toprule
    {\bf Attribute} & {\bf P} & {\bf B} & {\bf UU}\\ 
    \midrule
    \midrule
    {\em Demographics} & & & \\
    \small{Num. Workers} & 167 & 163 & 93 \\ %
    \small{Male} & 60.1\% & 68.7\% & 75.26\%\\ %
    \small{Female}  & 39.9\% & 31.3\% & 24.74\%\\ %
    \small{Average Age (in years)} & 37 & 38 & 37\\ 
    \midrule
    \midrule
    {\em Privacy Preference} &  &  & \\ 
    \small{Conscious (PC)} & 75.44\% & 78\%& 62.4\%\\ 
    \small{Not Conscious (NPC)} & 24.56\% & 22\%& 37.6\%\\ 
    \midrule
    \midrule
    {\em Education} & & & \\
    \small{Some High School} & 1.19\%& 1.22\%& 1.07\% \\
    \small{High School} & 11.97\%& 10.42\%& 1.07\%\\
    \small{Some College} & 17.96\%& 17.79\%& 5.37\%\\
    \small{Associate's} &10.17\%& 9.81\%& 3.22\%\\
    \small{Bachelor's} & 47.90\%& 47.85\%& 77.41\%\\
    \small{Graduate }& 10.77\%& 12.83\%& 11.82\%\\ 
    \bottomrule
    \end{tabular}
\end{center}
\caption{Demographics of participants of study reported in \S~\ref{sec:user}}
\label{table:demos}
\end{table}

\subsection{Adversarial Training: Additional Results}
\label{app:atraining}

\begin{table}[ht]
\begin{center}
  \begin{tabular}{ p{1.3cm} p{1.2cm} p{1.2cm} p{1.4cm} p{1.4cm}}
    \toprule
    \small{\bf Dataset} & \small{\bf Base ($\ell_2$)} & \small{\bf AT ($\ell_2$)} & \small{\bf Base (cos)} & \small{\bf AT (cos)}\\ 
    \midrule
    \midrule
    \small{LFW} & 79.9\% & 74.4\% & 75.8\% & 73.2\% \\
    \small{VGGFace2} & 83.6\% & 71.8\% & 82.4\% & 67.6\% \\
    \small{Celeb} & 85.7\% & 81.7\% & 80.4\% & 76.3\% \\
    \bottomrule
    \end{tabular}
\end{center}
\caption{Natural matching accuracies after adversarial training. Note that adversarial training decreases natural matching accuracy. As before, \textbf{Base} refers to the baseline natural accuracy (before adversarial training), and \textbf{AT} refers to the natural accuracy after adversarial training.}
\label{table:advmatch}
\end{table}

We choose to evaluate the results of adversarial training for both top-1 accuracy (Table~\ref{table:advtop}) and matching accuracy (Table~\ref{table:advmatch}). The top-1 case refers to the closest embedding from a bucket of labels whereas matching deals with the binary classification \ie match vs. mismatch. %

\subsection{Run-time}
\label{app:runtime}

We report micro-benchmarks related to run-time performance of our algorithms in Table~\ref{table:run-time}. Note that the $\ell_2$ variant of CW uses 8 binary search steps, and the $\ell_\infty$ variant of CW terminates at 10 trials.

\begin{table}[h]
\begin{center}
  \begin{tabular}{c c c c c}
    \toprule
    {\bf Attack} & {\bf Norm} &{\bf Model} & {\bf Avg. run-time (s)} & {\bf Batch Size}\\ 
    \midrule
    \midrule
    CW & ${2}$ & CSVC & 31.25 & 5 \\
    CW & ${2}$ & FLVT & 127.81 & 5 \\
    CW & ${\infty}$ & CSVC & 126.00 & 1 \\
    CW & ${\infty}$ & FLVT & 373.16 & 1 \\
    PGD & ${2}$ & CSVC & 6.40 & 5 \\
    PGD & ${2}$ & FLVT & 70.51 & 5 \\
    \bottomrule
    \end{tabular}
\end{center}
\caption{Run-time for mask generation. Each attack uses $N=100$ iterations. Run-times were evaluated on a server with 2 Titan XPs and 1 Quadro P6000. {\bf Model} refers to the surrogate model used to generate the masks.}
\label{table:run-time}
\end{table}

\subsection{Deployed Service}
\label{app:screenshots}

Figure~\ref{fig:website} contains screenshots from the service we deploy. A video highlighting its usability can be found here: \url{https://youtu.be/LJtcpZmz7JY}

\begin{figure*}[t]%
\centering
\includegraphics[width=\textwidth]{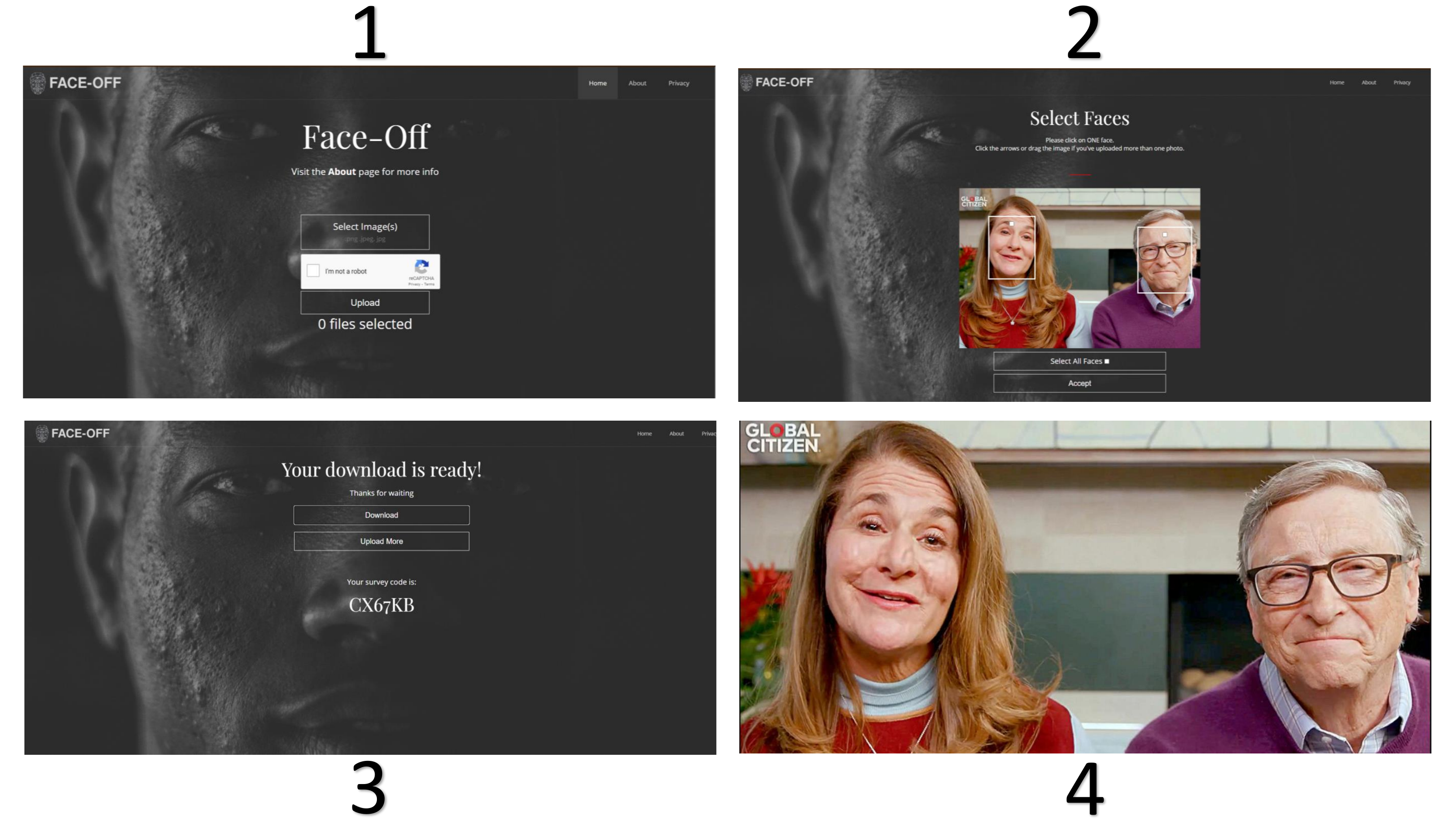}
\caption{Website view of \name's pipeline.}
\label{fig:website}
\end{figure*}
\subsection{Cropped Images}
\label{app:cropped}

\begin{figure*}[ht t]
\centering
	\begin{subfigure}{0.3\textwidth}
		\centering
		\includegraphics[width=0.85\textwidth]{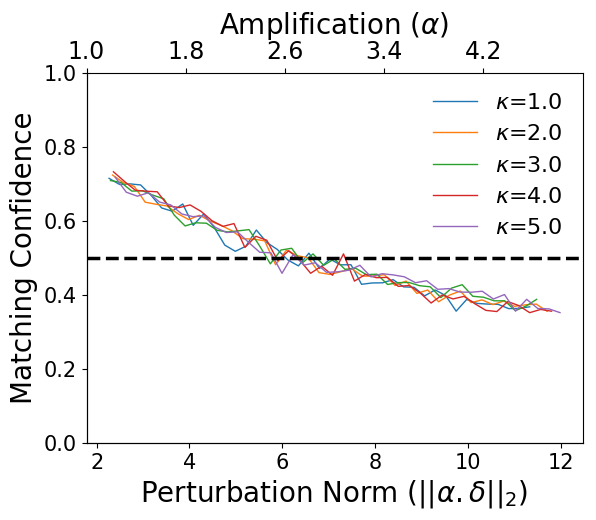}
        \caption{Azure}
        \vspace{0.1in}
		\label{fig:azure_cw_cc}%
	\end{subfigure}
    \begin{subfigure}{0.3\textwidth}
		\centering
		\includegraphics[width=0.85\textwidth]{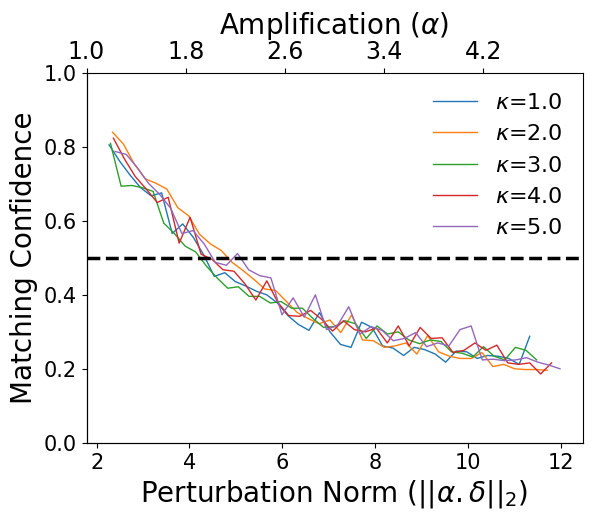}
		\caption{AWS Rekognition}
		\vspace{0.1in}
		\label{fig:awsverify_cw_cc}%
	\end{subfigure}
    \begin{subfigure}{0.3\textwidth}
		\centering
		\includegraphics[width=0.85\textwidth]{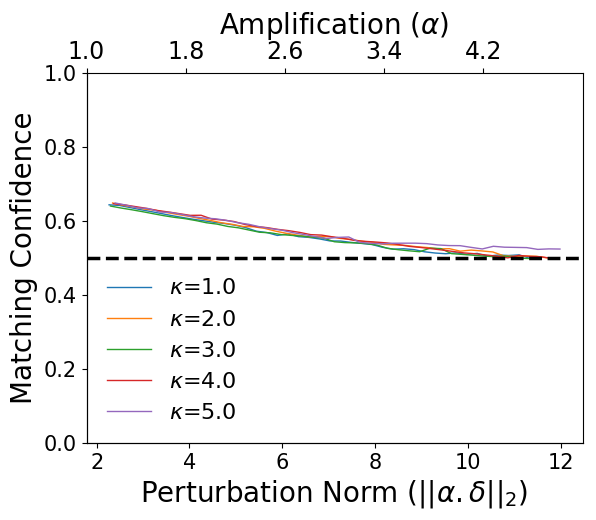}
		\caption{Face++}
		\vspace{0.1in}
		\label{fig:facepp_cw_cc}%
	\end{subfigure}
    \caption{2018: Transferability of cropped images generated using CW attack}
    \label{fig:cw_cc}
\end{figure*}

\begin{figure*}[]
\centering
	\begin{subfigure}{0.3\textwidth}
		\centering
		\includegraphics[width=0.85\textwidth]{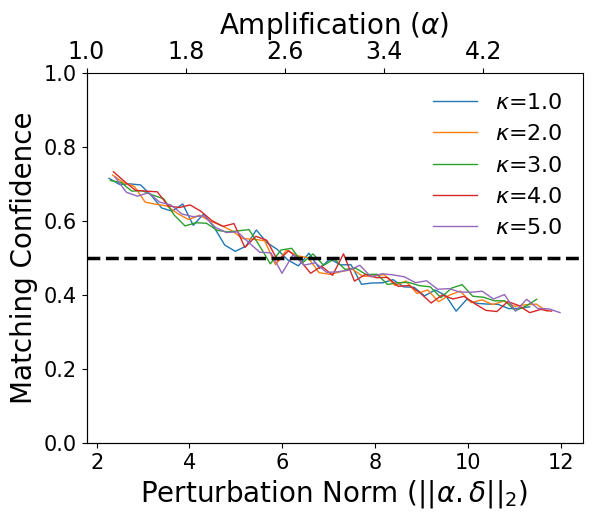}
        \caption{Azure}
        \vspace{0.1in}
		\label{fig:azure_cw_cc_2020}%
	\end{subfigure}
    \begin{subfigure}{0.3\textwidth}
		\centering
		\includegraphics[width=0.85\textwidth]{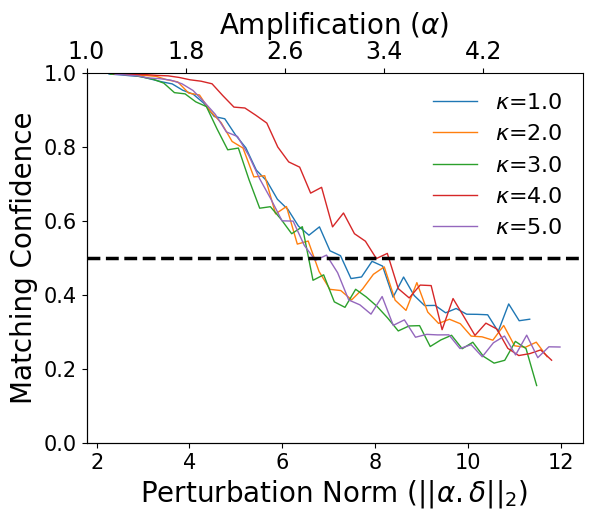}
		\caption{AWS Rekognition}
		\vspace{0.1in}
		\label{fig:awsverify_cw_cc_2020}%
	\end{subfigure}
    \begin{subfigure}{0.3\textwidth}
		\centering
		\includegraphics[width=0.85\textwidth]{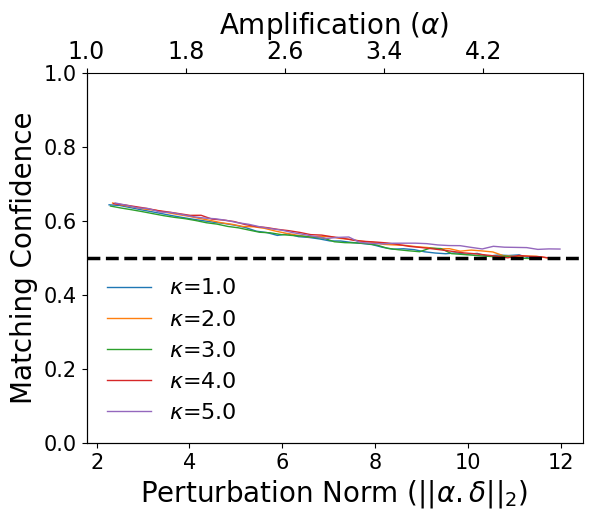}
		\caption{Face++}
		\vspace{0.1in}
		\label{fig:facepp_cw_cc_2020}%
	\end{subfigure}
    \caption{2020:Transferability of cropped images generated using CW attack}
    \label{fig:cw_cc_2020}
\end{figure*}

\begin{figure*}[]
\centering
	\begin{subfigure}{0.3\textwidth}
		\centering
		\includegraphics[width=0.85\textwidth]{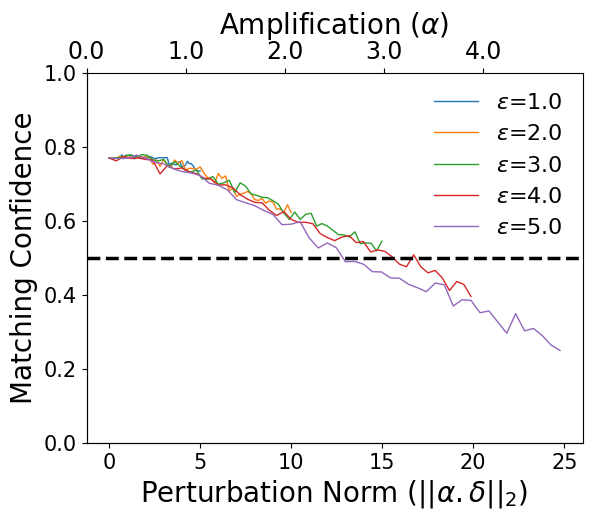}
        \caption{Azure}
        \vspace{0.1in}
		\label{fig:azure_pgd_cc}%
	\end{subfigure}
    \begin{subfigure}{0.3\textwidth}
		\centering
		\includegraphics[width=0.85\textwidth]{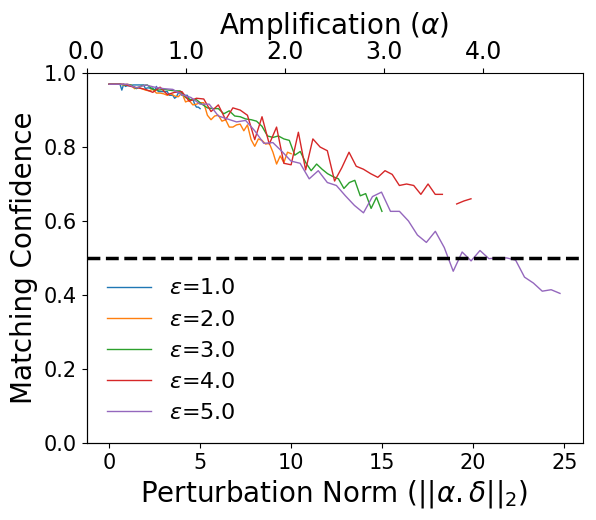}
		\caption{AWS Rekognition}
		\vspace{0.1in}
		\label{fig:awsverify_pgd_cc}%
	\end{subfigure}
    \begin{subfigure}{0.3\textwidth}
		\centering
		\includegraphics[width=0.85\textwidth]{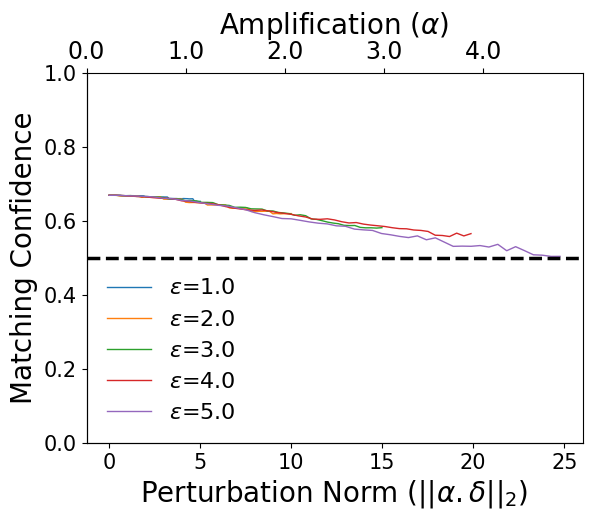}
		\caption{Face++}
		\vspace{0.1in}
		\label{fig:facepp_pgd_cc}%
	\end{subfigure}
    \caption{2018:Transferability of cropped images generated using PGD attack}
    \label{fig:pgd_cc}
\end{figure*}

\begin{figure*}[]
\centering
	\begin{subfigure}{0.3\textwidth}
		\centering
		\includegraphics[width=0.85\textwidth]{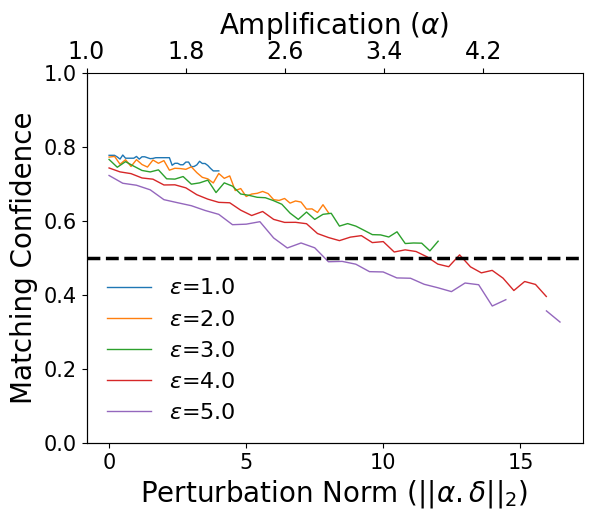}
        \caption{Azure}
        \vspace{0.1in}
		\label{fig:azure_pgd_cc_2020}%
	\end{subfigure}
    \begin{subfigure}{0.3\textwidth}
		\centering
		\includegraphics[width=0.85\textwidth]{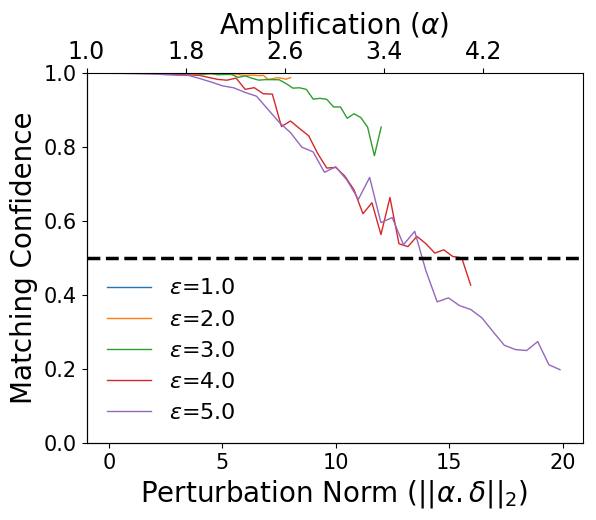}
		\caption{AWS Rekognition}
		\vspace{0.1in}
		\label{fig:awsverify_pgd_cc_2020}%
	\end{subfigure}
    \begin{subfigure}{0.3\textwidth}
		\centering
		\includegraphics[width=0.85\textwidth]{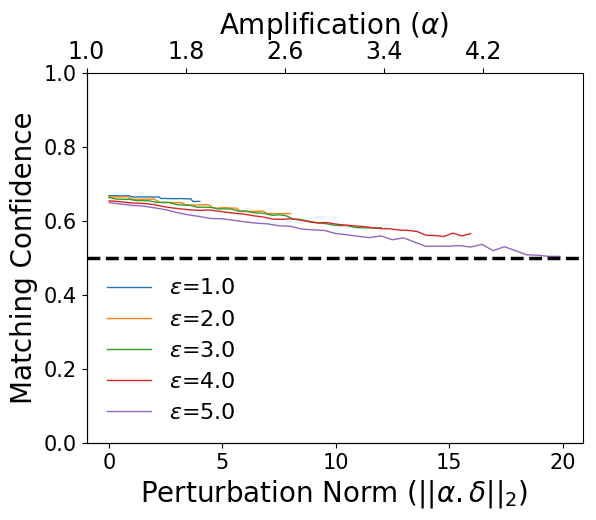}
		\caption{Face++}
		\vspace{0.1in}
		\label{fig:facepp_pgd_cc_2020}%
	\end{subfigure}
    \caption{2020: Transferability of cropped images generated using PGD attack}
    \label{fig:pgd_cc_2020}
\end{figure*}

In this experiment, we crop the adversarial inputs and compare it with a cropped reference (\ie an image with the true label of the cropped adversarial input). We observe that, similar to the uncropped images in \S~\ref{subsec:commercial} and~\ref{subsec:robustness}, the cropped images transfer to the black-box cloud APIs as well. This is the case for both CW (refer Figures~\ref{fig:cw_cc} and~\ref{fig:cw_cc_2020}) and PGD (refer Figures~\ref{fig:pgd_cc} and~\ref{fig:pgd_cc_2020}) attacks.

\end{document}